\documentclass[12pt]{iopart}
\usepackage[utf8]{inputenc}

\usepackage{subcaption}
\usepackage{graphicx}
\usepackage{amssymb}
\usepackage{enumerate}
\usepackage[gen]{eurosym}
\usepackage{hyperref}
\usepackage{xcolor}
\usepackage{siunitx}
\usepackage{tikz}
\usepackage{adjustbox}
\usepackage{braket}

\usetikzlibrary{patterns}
\usetikzlibrary{plotmarks}
\usetikzlibrary{patterns}

\usepackage[
backend=biber,
style=numeric,
eprint=true,
url=false,
sorting=none,
giveninits,
date=year,
isbn=false
]{biblatex}
\begin{document}

% ----
\title{Determining Absolute Neutrino Mass using Quantum Technologies}

\author{A~A~S~Amad$^1$, F~F~Deppisch$^2$, M~Fleck$^2$, J~Gallop$^3$, T~Goffrey$^4$, L~Hao$^3$, N~Higginbotham$^2$, S~D~Hogan$^2$, S~B~Jones$^{2\dag}$, L~Li$^1$, N~McConkey$^5$, V~Monachello$^2$, R~Nichol$^2$, J~A~Potter$^3$, Y~Ramachers$^4$, R~Saakyan$^{2*}$, E~Sedzielewski$^2$, D~Swinnock$^4$, D~Waters$^2$, S~Withington$^6$, S~Zhao$^6$ and J~Zou$^2$}
\address{$^1$ Department of Electronic and Electrical Engineering, Faculty of Science and Engineering, Swansea University, Swansea SA1 8EN, UK} 
\address{$^2$ Department of Physics and Astronomy, University College London, Gower Street, London WC1E 6BT, UK}
\address{$^3$ National Physical Laboratory, Hampton Rd., Teddington TW11 0LW, UK}
\address{$^4$ Department of Physics, University of Warwick, Coventry CV4 7AL, UK}
\address{$^5$ Particle Physics Research Centre, Queen Mary University of London, Mile End Road, London E1 4NS, UK}
\address{$^6$ Department of Physics, Clarendon Laboratory, Parks Rd, Oxford OX1 3PJ, UK}

\eads{$^{*}$\mailto{r.saakyan@ucl.ac.uk}, $^{\dag}$\mailto{seb.jones@ucl.ac.uk}}

% ling.hao@npl.co.uk
% john.gallop@npl.co.uk
% jamie.potter@npl.co.uk

% l.li@swansea.ac.uk 

% Daniel.Swinnock@warwick.ac.uk
% T.Goffrey.1@warwick.ac.uk
% y.a.ramachers@warwick.ac.uk

\vspace{10pt}
%\begin{indented}
%\item[]\today
%\end{indented}
% ----

\begin{abstract}
Next generation tritium decay experiments to determine the absolute neutrino mass require high-precision measurements of $\beta$-decay electron energies close to the kinematic end point. 
To achieve this, the development of high phase-space density sources of atomic tritium is required, along with the implementation of methods to control the motion of these atoms to allow extended observation times. 
A promising approach to efficiently and accurately measure the kinetic energies of individual $\beta$-decay electrons generated in these dilute atomic gases, is to determine the frequency of the cyclotron radiation they emit in a precisely characterised magnetic field. 
This cyclotron radiation emission spectroscopy technique can benefit from recent developments in quantum technologies. 
Absolute static-field magnetometry and electrometry, which is essential for the precise determination of the electron kinetic energies from the frequency of their emitted cyclotron radiation, can be performed using atoms in superpositions of circular Rydberg states.
Quantum-limited microwave amplifiers will allow precise cyclotron frequency measurements to be made with maximal signal-to-noise ratios and minimal observation times. 
Exploiting the opportunities offered by quantum technologies in these key areas, represents the core activity of the Quantum Technologies for Neutrino Mass project. 
Its goal is to develop a new experimental apparatus that can enable a determination of the absolute neutrino mass with a sensitivity on the order of $\qty{10}{\meV} / c^2$.

%\textcolor{red}{General comments/conventions:
%\begin{itemize}
%    \item Masses as meV$/c^2$ etc. If applied consistently then formulae should also be modified, e.g., Eq. (5). Do we use this convention only for numerical mass values?
%\end{itemize}
%}

\end{abstract}

%\submitto{\NJP}

\maketitle
\section{Introduction}
\label{sec:intro}

Seeking the origin of mass has been a driving force in the development of the Standard Model (SM) of particle physics. Due to the gauge-symmetric nature of the electromagnetic, weak and strong forces, fundamental particles would need to be massless if it were not for the Higgs mechanism of spontaneous symmetry breaking \cite{Englert:1964et, Higgs:1964pj, Guralnik:1964eu}. 
With the discovery of the Higgs boson at the Large Hadron Collider \cite{ATLAS:2012yve, CMS:2012qbp}, there is now a firm understanding of how quarks and charged leptons acquire mass, but this does not apply easily to neutrinos. Since only left-handed neutrinos interact via the weak force, and a hypothetical right-handed counterpart would not interact through the SM gauge forces at all, neutrinos had been considered massless in the SM. 
The left- and right-handed neutrino will interact with the Higgs via a Yukawa coupling (analogous to the other SM fermions), but probing it directly is very challenging, unless extended scenarios with the new sterile neutrino mass states are considered.
Direct experimental probes of neutrino masses, $m_i$, for example through tritium $\beta$-decay, seemingly confirmed this expectation, setting a stringent limit of $m_i < \mathcal{O}(1~\text{eV})$. This upper limit is already six orders of magnitude smaller than the mass scale of the other fermions: $m_i/m_e \lesssim 2\times 10^{-6}$, where $m_e$ is the mass of the electron, the next lightest fermion. 

The existence of neutrino oscillations, discovered by the Sudbury Neutrino Observatory~\cite{SNO2002} and Super-Kamiokande~\cite{SK1998} collaborations, now provides indirect but conclusive proof that at least two of the three SM neutrinos have nonzero masses, with the heaviest state weighing at least $m_\text{heaviest} \gtrsim \qty{50}{\meV}/c^2$, making the mass range from $\qty{50}{\meV}/c^2$ to $\qty{1}{\eV}/c^2$ a target for ongoing experimental studies.
% Mention SM massless neutrinos
Neutrino oscillations arise because of a misalignment between the charged-current neutrino interaction eigenstates $\nu_\alpha$, where $\alpha = e, \mu, \tau$, and the neutrino mass eigenstates $\nu_i$, with $i = 1, 2, 3$. These two basis sets are related by $\nu_\alpha = \sum_{i=1}^3 U_{\alpha i} \nu_i$, where $U_{\alpha i}$ is the Pontecorvo-Maki-Nakagawa-Sakata (PMNS) matrix, which describes neutrino flavour mixing.
Essentially all current experimental information on the PMNS matrix elements and the neutrino masses $m_i$ comes from observations of oscillations. These data place constraints on $U_{\alpha i}$, and the differences in the squares of the neutrino masses, $\Delta m^2_{ij} = m^2_i - m^2_j$. 
The parametrization of the PMNS matrix is not, however, unique; the standard form involves three mixing angles $\theta_{12}$, $\theta_{13}$, $\theta_{23}$ and a $CP$-violating phase $\delta_{CP}$~\cite{pdg2022}. 
The mass eigenstates are conventionally ordered by their electron-flavour content with $|U_{e1}|^2 > |U_{e2}|^2 > |U_{e3}|^2$: $\nu_1$ is the state most similar to $\nu_e$. 
The mixing angles and $\Delta m^2_{ij}$ are currently known to a precision of a few percent~\cite{nufit5.2}, but outstanding questions remain. 
These include the fact that whilst the sign of $\Delta m^2_{21}$ is known, the sign of $\Delta m^2_{32}$ is not, which leads to normal ordering (NO), in which $m_1 < m_2 < m_3$, and inverted ordering (IO), in which $m_3 < m_1 < m_2$. 
The $CP$-violating phase $\delta_{CP}$ also remains undetermined. 
Determining the mass ordering and measuring $\delta_{CP}$ are the primary physics goals of the future long-baseline neutrino oscillation experiments \cite{DUNE:2015lol, Hyper-KamiokandeProto-:2015xww}.

Although neutrino oscillation experiments have enabled numerous pivotal advances in neutrino physics, they provide no information about the absolute values of the masses. 
In this paper, we overview the Quantum Technologies for Neutrino Mass (QTNM) experiment, which aims to address this deficiency by contributing to the development of a new class of neutrino mass experiments. 
The technique, which goes under the generic name cyclotron radiation emission spectroscopy (CRES), is capable of measuring the effective $\beta$-decay neutrino mass $m_\beta$:
\begin{equation}
    m^2_\beta = \sum_{i=1}^3 \left|U_{ei}\right|^2 m_i^2 \,.
    \label{eq:mbeta}
\end{equation}
The long-term goal of QTNM, working with international colleagues, is to determine $m_\beta$, even if it is close to the minimally allowed value of 9~meV/$c^2$ set by neutrino oscillation observations. 
Our ambition is to exploit and further develop state-of-the-art experimental techniques that draw on recent advances in atomic, molecular, and optical physics and quantum microwave electronics and systems engineering. 
Our programme includes innovating, refining and increasing the technology readiness level of atomic quantum sensors for characterising magnetic and electric fields, and realising inward-looking arrays of quantum-noise-limited ($T_\text{noise} \lesssim$ 1~K at 18~GHz) coherent microwave receivers for observing large (1 m$^3$ to at least 10 m$^3$) volumes. 

Our article is organised as follows: In \sref{sec:neutrinomass}, previous work on determining absolute neutrino mass is summarised. This summary is followed, in \sref{sec:betadecay}, by a discussion of mass measurements based on the $\beta$-decay of radioactive isotopes, and the limitations imposed by existing experimental techniques in this field. In \sref{sec:CRES}, we focus on the method of CRES for determining precisely the kinetic energies of the radioactively liberated electrons, which in turn give information about the energies of the antineutrinos also released. 
In this context, the role of CRES in the next generation absolute neutrino mass experiments is described. 
Section \ref{sec:QTNM} summarises the theoretical, technological and instrument-definition studies underway in the UK, enabled through the QTNM partnership. 
More specifically, QTNM involves developing atomic tritium sources, and ensuring appropriate atomic quantum state selection and purity in the resulting gases: sections~\ref{sec:asource} and~\ref{sec:confinement}, respectively. 
CRES relies on the implementation of techniques for atomic magnetometry and electrometry to characterise precisely the primary static magnetic field that drives the cyclotron motion and that traps the electrons for a sufficient period of time that high spectral resolution is achieved: \sref{sec:magnetometry}. 
In \sref{sec:QTNMspec} an overview of the current conceptual design of the QTNM spectrometer is presented. 
This is followed in \sref{sec:Qamps} by a description of the quantum noise limited microwave amplifier technologies under development in the QTNM project. Finally, in \sref{sec:summary} conclusions are drawn.

%%%%%%%%%%%%%%%%%% Introduction %%%%%%%%%%%%%%%%%%
\section{Absolute neutrino mass}
\label{sec:neutrinomass}

A laboratory measurement of absolute neutrino mass is one of the most important experimental challenges facing the particle physics community. In fact, it is the only known particle whose mass has not yet been determined.
With oscillation experiments only sensitive to $\Delta m_{ij}^2$, alternative techniques are required to determine the individual masses. 
One approach to probing neutrino mass is based on the cosmological effect of the relic neutrino background on the Cosmic Microwave Background and the large-scale structure of the universe~\cite{eBOSS2021}. 
Such cosmic surveys are, however, only sensitive to the sum of the neutrino masses,
\begin{equation}
    \Sigma m_\nu = m_1 + m_2 + m_3 \,.
\end{equation}
The current limit is $\Sigma m_\nu < \qty{0.113}{\eV}/c^{2}$~\cite{DESI2024} with 90\% confidence, though it is affected by the choice of astrophysical data and depends on the neutrino mass ordering. 
It is weakened if an underlying cosmological model other than the standard minimal Lambda Cold Dark Matter ($\Lambda$CDM) model is used~\cite{RoyChoudhury:2019hls}. 
In particular, limits of $\Sigma m_\nu < \qty{280}{\meV}/c^2$ (NO) and $\Sigma m_\nu < \qty{290}{\meV}/c^2$ (IO) result (with 95\% confidence) from $\Lambda$CDM models with non-zero neutrino masses and free scaling of the so-called weak lensing amplitude $A_\text{lens}$ ($\Lambda$CDM + $\Sigma m_\nu$ + $A_\text{lens}$). 
While these cosmological constraints are stringent, they are not a substitute for laboratory measurements. 
At the same time, a laboratory measurement would provide a much needed prior constraint for cosmological fits. 
We see these various approaches as being highly complementary, rather than competitive.

\begin{figure}[t!]
    \centering
    \includegraphics[width=\textwidth]{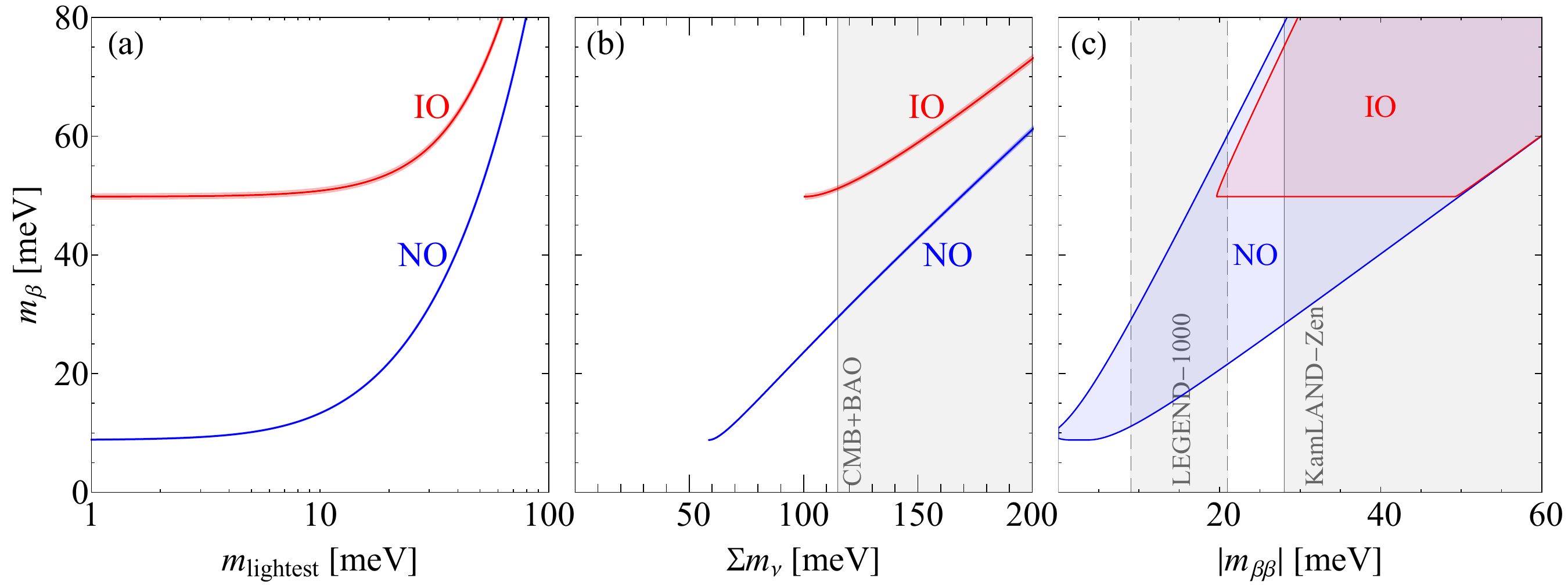}
    \caption{Dependence of the effective $\beta$-decay mass of the neutrino, $m_\beta$, on (a) the lightest neutrino mass $m_\text{lightest}$, (b) the sum of neutrino masses $\Sigma m_\nu$, and (c) the effective neutrinoless double $\beta$-decay mass $m_{\beta\beta}$. Neutrino oscillation parameters are taken from~\cite{nufit5.2} for NO (blue) and IO (red) with the bands indicating $1\sigma$ uncertainties.}
    \label{fig:mbb-vs-observables}
\end{figure}

Precise laboratory measurements of $\beta$-decay reactions in radioactive isotopes can be used to determine neutrino masses. 
In particular, the shape of the high-energy endpoint of $\beta$-decay electron spectra depends on $m_\beta$. Minimum values of $m_\beta$ are approached if the lightest neutrino is massless, i.e., $m_1 = 0$ for NO, or $m_3 = 0$ for IO. 
From neutrino oscillation data \cite{nufit5.2}, in the NO case the value is $m_\beta \geq (8.82\pm 0.11)~\text{meV}/c^2$ (NO), while for IO $m_\beta \geq 49.8^{+0.5}_{-0.4}~\text{meV}/c^2$. 
As can be seen in~\fref{fig:mbb-vs-observables}(a), the value of $m_\beta$ is strongly correlated with the lightest neutrino mass, $m_{\mathrm{lightest}}$. 
Measurements of $m_{\beta}$ therefore provide an important and very promising route toward a laboratory determination of the absolute mass scale. 
Several radioactive isotopes are suitable for experiments of this kind, with measurements using tritium currently playing a dominant role. 
The KATRIN experiment, which is based on a fundamentally different spectroscopic technique, provides the most stringent limit of $m_\beta < \qty{0.45}{\eV}/c^2$ at the 90\% confidence level~\cite{katrin2024}.

In determining the absolute neutrino mass scale, measurements of $m_\beta$ provide independent constraints on other physical processes. Firstly, neutrinoless double $\beta$-decay could occur if neutrinos are Majorana fermions, and the rate of this process is sensitive to the mass 
\begin{equation}
    \left|m_{\beta\beta}\right| = \left|
    \sum_{i=1}^3 \left|U_{ei}\right|^2 e^{2i\alpha_i} m_i \right| \,.\label{eq:mBB}
\end{equation}
In the standard parametrization, the complex phases are $\alpha_1 = 0$ and $0 \leq \alpha_{2,3} < \pi$, where $\alpha_{2,3}$ is unconstrained over their theoretically possible range: they cannot be probed in oscillation experiments.
The phases arise due to the Majorana nature of neutrinos, leading to a coherent sum over the mass eigenstates where cancellations are possible such that in the NO case $m_{\beta\beta} = 0$ may occur. 
In~\eref{eq:mBB}, we have explicitly extracted the Majorana phases $\alpha_i$ from the PMNS matrix elements to demonstrate this property.
This is possible for $9.1~\text{meV}/c^2 \lesssim m_\beta \lesssim 10.8$~meV$/c^2$. Currently, the most stringent bounds on $|m_{\beta\beta}|$ are set by the KamLAND-Zen collaboration, which constrained the neutrinoless double $\beta$-decay half life in ${}^{136}$Xe to $T_{1/2}^{0\nu\beta\beta}({}^{136}\text{Xe}) > 3.8\times 10^{26}$~yr with 90\% confidence~\cite{KamLAND-Zen:2024eml}. 
This corresponds to an upper limit $|m_{\beta\beta}| < 28 - 122$~meV/$c^2$, where the range indicates the theoretical uncertainty due to different nuclear structure models used to calculate the relevant nuclear matrix element. 
The planned future experiment LEGEND-1000 is designed to probe the neutrinoless double $\beta$-decay half life in $^{76}$Ge at the level of $T_{1/2}^{0\nu\beta\beta}({}^{76}\text{Ge}) \approx 10^{28}$~yr, corresponding to $|m_{\beta\beta}| < 9 - 21$~meV/$c^2$ \cite{LEGEND:2021bnm}.

In \fref{fig:mbb-vs-observables}, the relations between $m_\beta$ and the other neutrino mass parameters, $m_\text{lightest}$, $\Sigma m_\nu$ and $|m_{\beta\beta}|$, are shown. The curves presented are based on parameters obtained from a recent global fit of neutrino oscillation data for the cases of NO (blue) and IO (red) including $1\sigma$ uncertainties~\cite{nufit5.2}. 
In addition, in the case of $|m_{\beta\beta}|$, \fref{fig:mbb-vs-observables}(c), Majorana neutrinos are assumed and the Majorana phases are varied over their full range. 

The absolute mass scale of neutrinos, represented here by the lightest neutrino mass $m_\text{lightest}$, is not only a key parameter for neutrino physics, but also for particle physics more widely, and cosmology in general. 
Its determination would tell us the hierarchy of neutrino masses, $m_1 : m_2 : m_3$, and thereby test neutrino mass and fermion flavour models. 
For example, seesaw scenarios of mass generation near the electroweak scale incorporate right-handed neutrinos, which can also generate the correct matter-antimatter asymmetry of the Universe. 
Minimal versions of such scenarios predict a massless lightest neutrino and can thus be falsified if $m_\beta$ is measured above its smallest possible value~\cite{Deppisch:2015qwa}. 
The $\beta$-decay mass $m_\beta$, or, more generally, the tritium decay endpoint spectrum, appears as the most promising observable to determine the absolute neutrino mass scale, see \fref{fig:mbb-vs-observables}(a). 

Since the cosmological observable $\Sigma m_\nu$ depends on the specific cosmological model used, in contrast to a direct laboratory measurement, a concordant measurement of $\Sigma m_\nu$ and $m_\beta$ would provide a confirmation of the $\Lambda$CDM model, while measurements that deviate from the predicted relation in \fref{fig:mbb-vs-observables}(b) would indicate a departure from this scenario. 

Likewise, while neutrinoless double $\beta$-decay is crucial to understand the nature of neutrinos, its interpretation is considerably more model-dependent than tritium $\beta$-decay. As mentioned, it will only occur if neutrinos are Majorana fermions, whereas $m_\beta$ is essentially a kinematic parameter sensitive to neutrino masses irrespective of their nature. The neutrinoless double $\beta$-decay rate is also affected by other sources of lepton number violation in new physics scenarios beyond three active Majorana neutrinos. If $m_\beta$ is measured to be $m_\beta \gtrsim \qty{11}{\meV}/c^2$ this result would provide a lower limit for $|m_{\beta\beta}|$ as seen in \fref{fig:mbb-vs-observables}(c). Finally, for three active neutrinos, the mixing and mass differences obtained from oscillation experiments require that $m_\beta \geq (8.82\pm 0.11)$~meV$/c^2$. 
If a new measurement to this level of sensitivity does not yield a definite value, it would necessitate a rethinking of the nature of neutrinos and the source of their oscillations.

\section{Neutrino mass from $\beta$-decay}
\label{sec:betadecay}

% Description of how tritium decay spectrum depends on m_beta
Although laboratory measurements of $m_{\beta}$ can, in principle at least, be performed using a number of radioactive isotopes, the leading work on direct measurements of $\beta$-decay electron spectra has so far been carried out with tritium molecules, T$_2$, while complementary calorimetry experiments are carried out with $^{163}$Ho. 
However, to overcome challenges in accounting for energy imparted to rotational and vibrational degrees of freedom of the T$_2$ molecules, next generation experiments, such as those being developed by Project~8 and QTNM are moving towards the use of tritium (T) atoms.

The $\beta$-decay of a tritium atom $\mathrm{T} = {}^3\text{H}$ leads to the emission of an electron and an electron antineutrino, in the process producing a daughter helium-3 $\left(^3\text{He}\right)$ ion,
\begin{equation}
    \text{T} \to {}^3\text{He}^+ + e^- + \bar{\nu}_e. 
\end{equation}
The surplus energy, or $Q$-value, of the reaction is shared as kinetic energy between the decay products. 
Measurements of the mass difference between T and $^3$He$^+$ yield $Q = \qty{18575.72(7)}{\eV}$~\cite{tritiumQMeas}. 
The differential $\beta$-decay rate can be expressed in terms of the kinetic energy $E_e$ of the emitted electron \cite{OttenWeinheimer},
\begin{eqnarray}
    \frac{\rmd\Gamma}{\rmd E_e} &\approx& \frac{G_F \cos^2\theta_C}{2\pi^3} 
    (g_V^2 + 3g_A^2) F(2,E_e) |\mathbf{p}| (E_e + m_e) \nonumber \\
    &&\hspace*{0.2cm}\times \sum_{i=1}^3 \left|U_{ei}\right|^2 (E_0 - E_e) \sqrt{(E_0 - E_e)^2 - m_i^2}
    \,\Theta\left(E_0 - E_e - m_i\right) \,.
    \label{eq:fermiDecayRate}
\end{eqnarray}
Here, $G_F = 1.17\times 10^{-5}$~GeV$^{-2}$ is the Fermi coupling constant, $\theta_C = 0.974$ the Cabibbo angle \cite{pdg2022}, and $g_V = 1$ and $g_A \approx 1.247$ \cite{Simkovic:2007yi} are the nucleon vector and axial couplings, respectively. 
$\Theta$ is the Heaviside step function.

The Fermi function $F(Z, E_e)$ for He (nuclear charge $Z = 2$) in this expression describes the final state interactions of the emitted electron. These include the Coulomb interaction of the $^3$He nucleus with the $\beta$-decay electron with corrections including photon loops, soft photon emission and nuclear recoil applied. The three-momentum of the electron is $|\mathbf{p}| = \sqrt{E_e(E_e+2m_e)}$ and the endpoint $E_0$ of the electron spectrum is $E_0 = Q - E_\text{rec}$, where $E_\text{rec}$ is the recoil energy of the daughter nucleus. 

\begin{figure}[t!]
    \centering
    \includegraphics[width=0.49\textwidth]{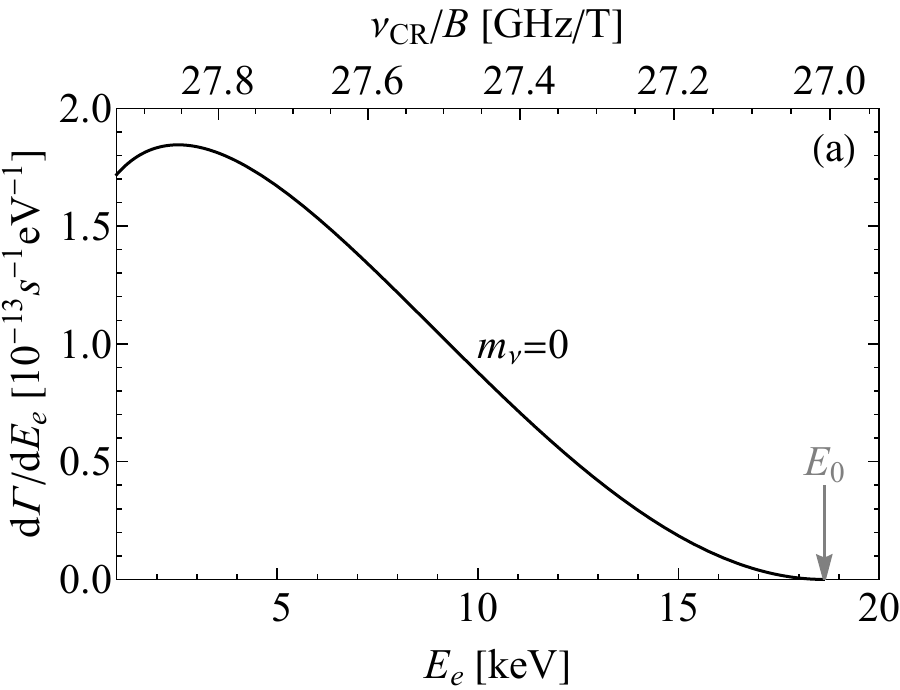}
    \includegraphics[width=0.49\textwidth]{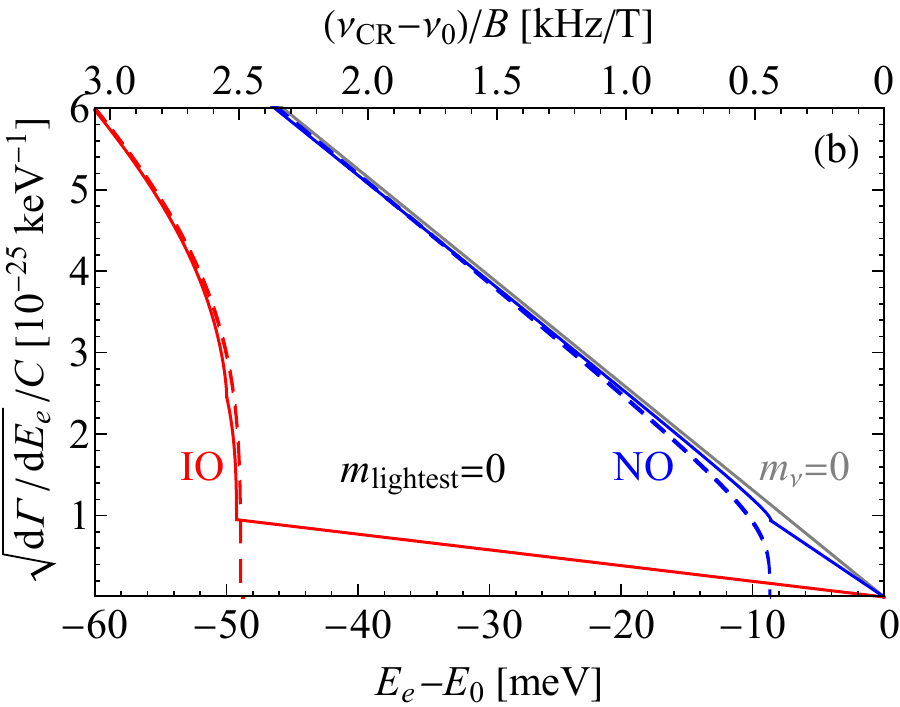}
    \caption{(a) Tritium decay spectrum $\rmd\Gamma/\rmd E_e$ with respect to the emitted electron kinetic energy $E_e$. The solid black curve ($m_\nu = 0$) indicates the spectrum for a single, massless antineutrino.
    (b) Kurie plot of $C^{-1}\sqrt{\rmd\Gamma/\rmd E_e}$ near the endpoint for a single massless electron-antineutrino (continuous grey curve), and three active neutrinos in NO (continuous blue curve) and IO (continuous red curve) with a massless lightest neutrino. The dashed curves in (b) represent spectra in which the three neutrino mass contributions are replaced by $m_\beta$.}
    \label{fig:spec}
\end{figure}
Equation~\eref{eq:fermiDecayRate} represents a sum over the contributions from each of the neutrino mass eigenstates, $m_i$, weighted by $|U_{ei}|^2$. 
This reflects the incoherent nature of the emission of the three neutrino-mass eigenstates, which is allowed if kinematically possible. 
The differential $\beta$-decay rate for T atoms is presented in \fref{fig:spec}. 
The full $\beta$-decay spectrum is shown in panel (a), while panel~(b) shows an expanded Kurie plot at energies close to the endpoint. Neutrino masses below 1~eV$/c^2$, as required to satisfy current experimental constraints, only affect the $\beta$-decay electron kinetic energy distribution very close to the endpoint. In this region, almost all the surplus energy of the reaction is carried away by the electron. The calculated spectra in \fref{fig:spec}(b) show the clear differences in the spectra in the cases of NO (continuous blue curve) and IO (continuous red curve) when the lightest neutrino is massless. The spectrum associated with a massless electron antineutrino is given by the continuous grey curve. The dashed curves in this panel show the spectra obtained by replacing the three neutrino mass contributions with the corresponding value of $m_{\beta}$ such that $m^2_\beta = \sum |U_{ei}|^2 m_i^2$, and 
\begin{equation}
    \frac{\rmd\Gamma}{\rmd E_e} \approx C (E_0 - E_e) 
    \sqrt{(E_0 - E_e)^2 - m_\beta^2} \, 
    \Theta\left(E_0 - E_e - m_\beta\right) \,,
    \label{eq:fermiDecayRateEffective}
\end{equation}
where $C$ is a normalisation factor that incorporates the constant, or slowly varying, terms near the endpoint, in \eref{eq:fermiDecayRate}. As $m_\beta$ increases, the endpoint in the electron spectrum moves to lower energies. 
While the resulting spectra are seemingly very different, \eref{eq:fermiDecayRateEffective} is a very good approximation when considering the integrated rate in the range $\Delta E_e \gg m_\beta\,c^2$. 

%% Brief overview of the technologies

% MAC-E
Currently, the tightest bounds on $m_{\beta}$ have been determined by the KATRIN experiment~\cite{katrin2021}. 
This experiment measures the $\beta$-decay electron spectrum of T$_2$. 
This is done using a magnetic adiabatic collimation -- electrostatic (MAC-E) filter. The MAC-E technique involves adiabatic collimation and transport of $\beta$-decay electrons from a region of high magnetic field in which they are produced to an integrating electrostatic spectrometer, where the electron flux is measured for a selected retarding potential. 
A complete electron spectrum is obtained by operating the spectrometer sequentially over a range of retarding potentials. 
This approach was pioneered in experiments in Mainz~\cite{mainzExp} and Troitsk~\cite{lobashev2003} which placed upper bounds of $m_{\beta} < \qty{2.05}{\eV}/c^{2}$~\cite{troitsk2011}, and $m_{\beta} < \qty{2.3}{\eV}/c^{2}$~\cite{mainz2005} with 95\% confidence, respectively. 
The tightest upper bound on $m_{\beta}$ currently set by KATRIN is $m_{\beta} < \qty{0.45}{\eV}/c^{2}$ with 90\% confidence~\cite{katrin2024}. 
There is scope for measurements of $m_{\beta}$ with an improved precision, with KATRIN's final sensitivity estimated to be better than $\qty{300}{\meV} / c^{2}$~\cite{katrin2024}. 
However, to go beyond this, scaling limitations in the MAC-E spectrometry technique impose experimental constraints that are challenging to overcome without significantly increasing the size of an already large instrument. 
The energy resolution of the MAC-E filter in KATRIN is limited by the spectrometer size, scaling as $(R_\text{sc}/R_\text{spec})^2$, where $R_\text{sc}$ and $R_\text{spec}$ are the radii of the tritium source and the spectrometer, respectively.
The KATRIN spectrometer, with a diameter of \qty{9.8}{\m} and a length of \qty{23.3}{\m} operating at \qty{d-11}{\milli\bar}~\cite{katrin2021}, is already the largest ultra-high-vacuum vessel in the world.
Therefore, KATRIN may be anticipated to be the ultimate implementation of the MAC-E filter technique for absolute neutrino mass measurement.
Furthermore, when studying the $\beta$-decay process using molecules, some of the surplus energy of the reaction can be imparted to the motion of the nuclei, through rotational or vibrational excitation.
Since there are many rotational and vibrational states available, it is difficult in any individual $\beta$-decay event to determine the energy lost to these degrees of freedom. 
This complication ultimately limits the precision with which $m_\beta$ can be determined from the $\beta$-decay electron spectrum of molecular tritium~\cite{molTEffects}.

% Calorimetry
Calorimetry experiments that complement direct measurements of $\beta$-decay electron spectra are currently performed with $^{163}\text{Ho}$~\cite{Ho163Original}. 
This decays by atomic electron capture in the reaction
\begin{equation}
    ^{163}\text{Ho} + e^{-} \rightarrow ^{163}\text{Dy} + \nu_{e} \, .
\end{equation}
This type of reaction can be exploited to determine $m_{\beta}$ through microcalorimetry if all of the surplus energy from the reaction is converted into heat in the calorimeter. 
In a similar way to the $\beta$-decay-electron spectrum, the neutrino mass affects the shape of the decay spectrum close to the endpoint.
Currently, the Electron Capture $^{163}\text{Ho}$ experiment (ECHo)~\cite{ECHoExp2014, ECHoExp2017}, and the HOLMium Experiment for neutrino mass Search (HOLMES)~\cite{HOLMESExp} collaborations are pursuing this approach, each with different sensor technologies.
The ECHo collaboration uses magnetic metallic calorimeters~\cite{MMCs} to measure the thermal energy, while the HOLMES experiment exploits superconducting transition-edge sensors~\cite{holmesTES}. 
However, in these experiments, it is necessary to trade off high decay rates, and the statistical advantage they provide, with energy resolution and event pileup due to the necessarily slow response of large-volume calorimeters. 
This is achieved through the use of a large number of microcalorimeters, rather than just one large calorimeter. 
Nevertheless, challenges remain in using these techniques to reach the precision needed for a definitive measurement of $m_\beta$.

In general, the bounds on $m_\beta$, and hence the absolute neutrino mass, determined through laboratory experiments are limited at present by two main factors: (1) challenges in implementing electron spectrometers based on MAC-E filters with higher energy resolution, and (2) the use of T$_2$ molecules as the $\beta$-decay electron source. 
To move beyond these, new experimental approaches to electron spectrometry are required that can be implemented in an apparatus suitable for measuring the $\beta$-decay electron spectra of atoms. 

\section{Cyclotron Radiation Emission Spectroscopy}
\label{sec:CRES}

The above considerations led to the proposal by Monreal and Formaggio~\cite{MonrealFormaggio2009} to move to a new approach for  measuring the kinetic energies of $\beta$-decay electrons in atomic tritium experiments. 
This is based around measuring the frequency of the cyclotron radiation emitted when an electron orbits in a strong static magnetic field. 
For fields of 0.5-1~T, the lowest-order tone of the emitted radiation lies in the 10-30~GHz range, and so microwave instrumentation can be used to measure the frequency, and hence energy, exceedingly accurately if the static magnetic field is homogeneous and known to high precision. 
This technique, which is known as Cyclotron Radiation Emission Spectroscopy, lies at the core of a new generation of absolute neutrino mass experiments, Project~8 and QTNM.
The He6-CRES collaboration has also demonstrated the use of CRES to detect electrons and positrons, generated by the $\beta$-decay of $^{6}$He and $^{19}$Ne, respectively, with energies of up to \qty{2.1}{\MeV}~\cite{Byron2023}.
It should also be emphasised that CRES is also a key part~\cite{PTOLEMY_CRES} of the PTOLEMY~\cite{PTOLEMY2019} experiment's plan to measure relic neutrinos.
A recent review of the technique may be found in~\cite{OblathVanDevender2024}.

Measurements of cyclotron frequencies of low-energy leptons have played a central role over the last half a century in precision tests of fundamental physics, including, for example, the determination of the electron and positron $g$ factors~\cite{graeff69a, schwinberg81a, brown86a, fan23a}. 
In addition, cyclotron radiation cooling of low-energy electron and positron plasmas in strong magnetic fields is exploited for applications in positronium and antihydrogen production~\cite{Amoretti2002}. 
For absolute neutrino mass measurements, however, the key objective is to measure the energies of single electrons at the point of their liberation with minimal disturbance. 
From this perspective, single-electron CRES represents an excellent opportunity and a fascinating problem. 
It can be implemented over a wide range of kinetic energies, including those around the end point of the $\beta$-decay spectrum of tritium at 18.6~keV, but it is demanding to achieve high precision over volumes of 1-10 m$^3$. 
These challenges are described in detail in the following.

For light particles, such as electrons and positrons, relativistic corrections play a key role in determining the frequency of the cyclotron radiation emitted when measured in the laboratory frame of reference. 
In this situation, the cyclotron radiation frequency $\nu_{\mathrm{CR}}$, of an electron with kinetic energy $E_e$ undergoing cyclotron motion in a magnetic field $B$ can be expressed as~\cite{MonrealFormaggio2009}
\begin{equation}
    \nu_{\mathrm{CR}} = \frac{e B}{2\pi(m_e + E_e/c^2)},
\label{eq:fCR}
\end{equation}
where $e$ and $m_e$ are the elementary charge and the electron mass, respectively, and $c$ is the speed of light in vacuum. 
The dependence on the kinetic energy in the denominator in \eref{eq:fCR} reduces the cyclotron frequency measured in the laboratory frame of reference for larger values of $E_e$.

The CRES technique can be implemented to collect cyclotron radiation from $\beta$-decay electrons generated in dilute atomic sources over extended volumes. 
This can be done by enclosing the measurement region in the CRES spectrometer in a waveguide or cavity, or surrounding it by arrays of antennas. 
Because the cyclotron radiation field has a well-defined spatial distribution, these components can be optimised for the frequency of interest to maximise the collection efficiency. 
This, together with standard approaches to microwave frequency metrology, allows for the precise determination of the cyclotron frequency of single electrons.

However, the opportunity to make use of spatially extended atomic sources of $\beta$-decay electrons in a CRES spectrometer also brings with it challenges. 
To precisely determine electron kinetic energies from the frequency of the emitted cyclotron radiation, it is essential to precisely map the magnetic field distribution  throughout the whole of the measurement region. 
This is particularly demanding for a large-volume field of view, operated at temperatures below \qty{4}{\K}, which is necessary to minimise thermal noise. 
In addition to the precise characterisation of the static magnetic field, it is also necessary to characterise, and minimise stray electric fields that arise from patch potentials, adsorbates or imperfections on the cold interior surfaces of the instrumented region. 

If the CRES technique is used to measure $\beta$-decay electron energies close to the atomic tritium end point at 18.6~keV,  the frequency of the cyclotron radiation emitted in a magnetic field of 0.1~T (1~T) is $\nu_{\mathrm{CR}} = 2.7$~GHz (27~GHz). 
Differentiating \eref{eq:fCR} then gives,
\begin{equation}
	\frac{\rmd E_e}{E_e} = 
	-\left(1 + \frac{m_e c^2}{E_e}\right) 
	\frac{\rmd \nu_\text{CR}}{\nu_\text{CR}} \, ,
	\label{eq:fCR_diff}
\end{equation}
where, in this case, $1 + m_e c^2 / E_e \approx 28$. 
Consequently, to determine an electron kinetic energy to a precision of $\pm100$~meV from the frequency, $\nu_{\mathrm{CR}}$ must be measured to a precision of $\simeq \pm\qty{500}{\Hz}$ ($\pm\qty{5}{\kHz}$), and the magnetic field strength at the position of the radiating electron must be known to a precision of $\sigma_B \simeq \pm 20~$nT ($\pm200$~nT).

The use of the CRES technique to detect and measure the kinetic energies of single electrons was first demonstrated by the Project 8 collaboration in experiments with radioactive $^{83\,\mathrm{m}}$Kr~\cite{Project8_2015}. 
Upon internal conversion, this isomer of $^{83}$Kr emits electrons in several comparatively narrow energy bands. 
Those with energies close to 30.4~keV and 17.8~keV were detected by observing the cyclotron radiation emitted at frequencies between 25 and 26~GHz in a magnetic field of 0.95~T. 
The cyclotron radiation was collected in a waveguide before being amplified using cryogenic HEMT amplifiers operating at 50~K. 
These experiments allowed the change in the cyclotron frequency to be observed as the electrons radiated, and lost kinetic energy, over time scales of milliseconds. 
Most recently, this work was extended to perform a first measurement of the $\beta$-decay electron spectrum of T$_2$ by CRES~\cite{Project8_2023}. 
This pioneering result yielded a FWHM electron-kinetic-energy resolution of 1.66~eV, and paves the way for future experiments directed toward the determination of electron kinetic energies to a precision of $\pm100$~meV and below.

\section{The QTNM Project}
\label{sec:QTNM}

\subsection{General considerations}
\label{sec:qtnm:goals}

The QTNM project aims to determine the absolute neutrino mass from the $\beta$-decay electron spectrum of T atoms.
This will be achieved by building on the pioneering work of the Project 8 collaboration and their use of CRES to measure kinetic energies of individual $\beta$-decay electrons close to 18.6~keV, while also taking advantage of state-of-the-art techniques from experimental Atomic, Molecular and Optical (AMO) physics and recent advances in the development of ultra-low-noise microwave amplifiers and other quantum technologies.
The ultimate goal of QTNM is to develop a scalable experimental apparatus with a sensitivity that could allow access to values of $m_{\beta}$ approaching the lower bound of $\sim$10~meV/$c^2$ set by observations of neutrino oscillations.
Beyond this, efforts will also be made to extend the operation of the QTNM CRES electron spectrometer to record $\beta$-decay-electron spectra for T atoms over an extended energy range and open opportunities for model-independent searches for sterile neutrinos, and exotic interactions of active or sterile neutrinos~\cite{canning2023}.

The sensitivity of an atomic tritium CRES measurement of $m_{\beta}$ can be estimated by using the analytic approach described in \cite{nuMassReview} and \cite{Doe2013}. 
This sensitivity is considered to be the standard deviation $\sigma_{m^2_{\beta}}$, on a measurement of $m^2_\beta$.
It can therefore be expressed in terms of a 90\% confidence limit (CL) on $m_\beta$ as $\sqrt{1.28 \sigma_{m^2_{\beta}}}$, and depends on the number of electrons detected in the energy region of interest close to the end-point of the $\beta$-decay spectrum, and the background rate. The main additional parameters of a near-ideal experiment that affect the sensitivity, can be inferred from \eref{eq:fCR} to be: (i) the precision with which the cyclotron frequency can be determined; and (ii) the accuracy with which the static magnetic field can be mapped and corrected for.

In general, the precision with which the frequency of a time-periodic signal can be determined depends on the observation time $t_{\text{obs}}$. 
For a deterministic signal in noise, the Cram\'er-Rao bound~\cite{Cramer1946, Rao1992} provides an upper limit on the precision with which an unbiased estimator of a signal parameter can be determined for a given value of $t_{\text{obs}}$. 
Another useful limit is simply the inverse of the observation time: $t_{\text{obs}}^{-1}$.
In the case of the Cram\'er-Rao bound for a sine wave with a linear frequency chirp, sampled at a rate $f_s$, the standard deviation on the initial frequency is
\begin{equation}
	\mathrm{std} \left(\hat{\nu}_0\right) \gtrsim \frac{4\sigma_{\mathrm{noise}}}{\pi} \sqrt{\frac{3}{P_\text{sig} t_\text{obs}^3 f_s}} \, ,
	\label{eq:fVar}
\end{equation}
where $\sigma^2_{\mathrm{noise}}$ is the variance of the background white Gaussian noise, and $P_\text{sig}$ is signal power. 
Although this expression contains several assumptions, it indicates the key parameters that determine the precision with which the initial cyclotron frequency is recovered: (i) the effective noise temperature of the heterodyne receiver, (ii) the sampling rate, and (iii) the observation time.

% ------------------------------------------------------------
% ESTIMATED SENSITIVITIES - FIGURE AND EXPLAINATION
% ------------------------------------------------------------

\begin{figure}[t!]
	\centering	
	\includegraphics[width=0.8\linewidth]{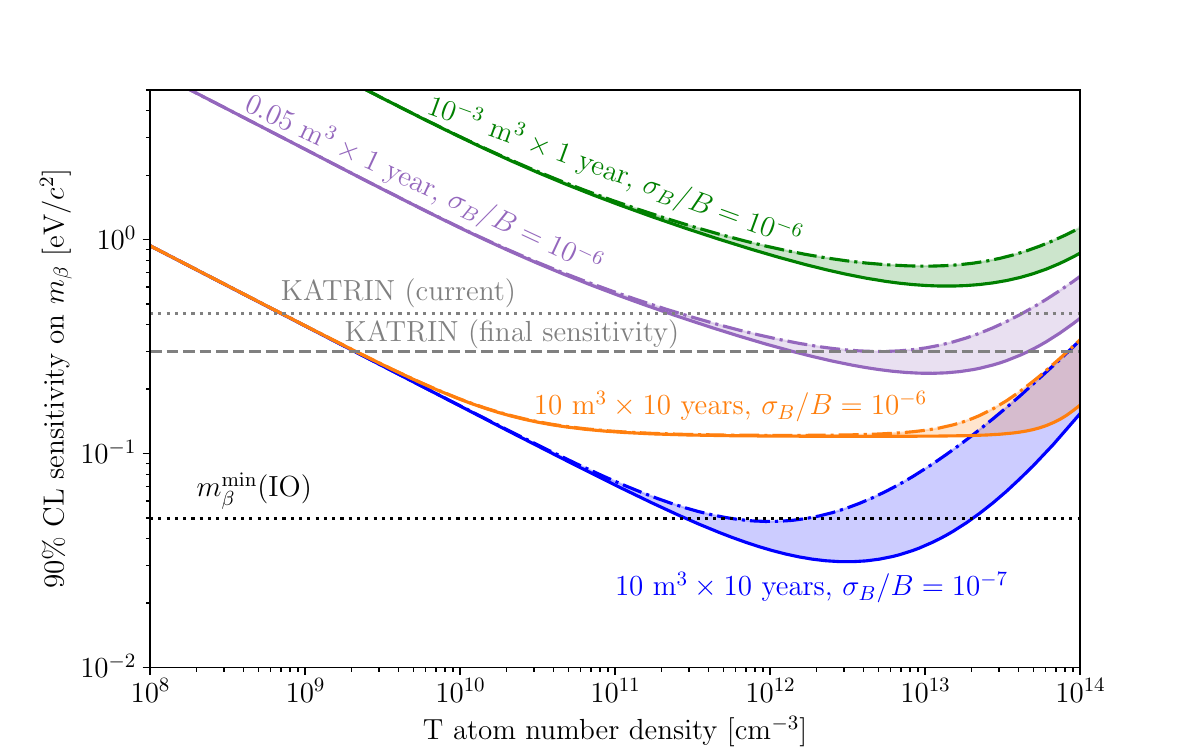}
	\caption{
		Dependence of the sensitivity of a CRES experiment to measure $m_\beta$, on the T atom number density in the measurement volume. 
  The set of volumes considered are indicated. In each case, the coloured bands represent the range of sensitivities expected from a frequency measurement precision set by the Cram\'er-Rao bound (equation~\ref{eq:fVar}, continuous curve) to that set by $t_\text{obs}^{-1}$ (dashed-dotted curve). 
  In all cases, the magnetic field strength was chosen to be 1~T (see text for details).
  Horizontal lines denoting the current, and ultimate projected limit of the KATRIN experiment with T$_2$ molecules~\cite{katrin2024}, and the minimally allowed value of $m_\beta$ for IO neutrinos, i.e., $m_\beta^{\mathrm{min}}(\mathrm{IO})$, are also shown. 
   }
	\label{fig:mb}
\end{figure} 

The estimated sensitivities of a T atom CRES experiment to the value of $m_{\beta}$ are shown in \fref{fig:mb}. 
In this figure, instrumented CRES measurement regions with a range of different volumes, from $10^{-3}$~m$^{3}$ to $10$~m$^{3}$, and in which the magnetic field strength is set to be 1~T are considered. 
For each of these, and a measurement-campaign duration of 1 or 10 years, the effect of the T atom number density on the precision with which $m_{\beta}$ can be recovered is presented. 
In each case, a shaded band is shown, which is bounded by two lines: a continuous curve corresponding to the precision with which the initial cyclotron frequency can be determined according to the Cram\'er-Rao bound using an observation time $t_{\text{obs}}$, and the dashed-dotted curve corresponding to the limit set by the inverse of $t_{\mathrm{obs}}$. 
For all calculations, a density- and energy-independent background rate is included, which scales with the volume. 
For an instrumented volume of $10^{-3}$~m$^3$, this rate is $10^{-9}~\text{s}^{-1}~\text{eV}^{-1}$ which is consistent with limits set by recent low-volume CRES test experiments with T$_2$~\cite{Project8_2023, Project8_2024}. 

At low number densities in \fref{fig:mb}, the sensitivity to $m_{\beta}$ is limited by the $\beta$-decay-electron count rate. 
Consequently, as the T atom number density rises from $10^8$ to $10^{11}$~cm$^{-3}$ the sensitivity also increases because of the commensurate increase in $\beta$-decay events. 
For atom number densities above $10^{11}$~cm$^{-3}$, the rate at which a radiating electron scatters from the background gas of neutral T atoms plays an increasingly significant role, because of the limit that this scattering process imposes on the maximal value of $t_{\mathrm{obs}}$. In this regime, the precision with which the cyclotron frequency can be recovered according to the Cram\'er-Rao bound offers a higher sensitivity to $m_{\beta}$, for any given atom number density, than the precision set by $t_{\mathrm{obs}}^{-1}$. 
At densities above $\sim10^{13}$~cm$^{-3}$, the scattering rate tightly restricts, and ultimately leads to a reduction in, the achievable sensitivity. 

The upper three curves (green, violet and orange) in \fref{fig:mb} are the results of calculations in which the fractional uncertainty in the strength of the magnetic field experienced by the radiating electron is $\sigma_B/B = 10^{-6}$ such that $\sigma_B = 1~\mu$T in the case considered. 
Under these conditions, a compact CRES apparatus with an instrumented volume of $10^{-3}$~m$^3$ is expected to provide a sensitivity below 1~eV/$c^2$ for T atom number densities between $10^{12}$ and $10^{13}$~cm$^{-3}$. 
Bearing in mind the characteristics and specifications of currently available H atom sources that could be operated to generate T atoms, and superconducting solenoid magnets, a demonstrator experiment of this scale represents an accessible first phase T-atom CRES prototype.
This would not result in a sensitivity to $m_{\beta}$ below the current limit set by the KATRIN collaboration in their work with T$_2$ (upper dotted horizontal line). 
It would, however, pave the way towards a measurement sensitivity beyond that expected from the final limit projected for KATRIN~\cite{katrin2024} (dashed horizontal line), if the measurement volume is increased to 0.05~m$^3$ (violet curves). 

If the CRES measurement volume can be increased to 10~m$^3$, in an ultimate large scale facility, the sensitivity to $m_{\beta}$ will no longer be limited by the count rate or cyclotron radiation observation time, but will become dominated by the accuracy with which the magnetic field experienced by each $\beta$-decay electron is known. 
If this magnetic field is characterised with an accuracy of $\sigma_B/B = 10^{-6}$ and a 10-year-long measurement campaign is undertaken, the lower bound on the achievable sensitivity approaches $100$~meV$/c^2$ (orange curves in \fref{fig:mb}). 
This sensitivity is not expected to be strongly dependent on the T atom number density within the range from $10^{11}$ to $10^{13}$~cm$^{-3}$, and would be an improvement on the projected final limit of the KATRIN experiment. 
Indeed, it would also bring a CRES experiment with T atoms into a regime in which the sensitivity to $m_{\beta}$ begins to surpass the 100~meV$/c^2$ limit of experiments with T$_2$ that arise because of the uncertainty in the internal molecular rotational and vibrational state distributions after $\beta$ decay. 
However, it will not permit access, for example, to the minimal value of $m_{\beta}^{\mathrm{min}}(\mathrm{IO})\simeq49.8$~meV$/c^2$ expected in the IO scenario (lower dotted horizontal line). 
To enhance the sensitivity of a T-atom CRES experiment toward values of $m_{\beta}$ in this range and below, it is necessary to improve the magnetic field characterisation. 
This will be challenging in a facility with a total instrumented volume of 10~m$^3$. 
However, if an order of magnitude improvement can be achieved, so that $\sigma_B/B = 10^{-7}$, a sensitivity to values of $m_{\beta}\simeq30$~meV$/c^2$ could become accessible. 
In addition to complete coverage of the range of values of $m_{\beta}$ associated with the IO scenario, this would also allow coverage of a significant portion of the NO parameter space. 
Further improvements in sensitivity toward the $\sim\qty{10}{\meV}/c^2$ lower bound on $m_{\beta}$ for the case of normal ordering and $m_1 = 0$, will be dominated by the requirement to significantly increase the instrumented volume with a smaller increase in magnetic field accuracy also necessary. 

The above analysis is carried out under the assumptions that the T atoms are stationary at the time of $\beta$-decay, and no T$_2$ molecules are present as contaminants in the CRES volume. 
The first of these holds for all scenarios considered, provided the atoms move at speeds below $\sim300$~m/s. In the three lower sensitivity scenarios (the upper three sets of green, purple and orange curves), it is expected that the effect of the motion of the atoms can be neglected up to speeds of $\sim1000$~m/s. 

The assumption that no significant contamination of the CRES volume with T$_2$ applies provided the molecule:atom ratio is below $10^{-5}$ and contributions to the electron spectrum from vibrational and rotational excitation after $\beta$-decay are minimal. 
These conditions are compatible with the phase-space characteristics of cryogenic sources of H and D atoms that could be operated to produce T atoms, and the possibility to separate paramagnetic T atoms from T$_2$ molecules, which do not possess a significant magnetic dipole moment, using inhomogeneous magnetic fields. 
The analysis assumes that 20\% of all $\beta$-decay electrons produced within the instrumented volume are detected, before accounting for efficiency losses due to scattering, and 10\% of the radiated power from each individual electron is collected and subsequently digitised at a rate of $f_{\mathrm{s}}=1$~GHz. 
The background microwave noise temperature was considered to be $T_{\mathrm{noise}} = 5$~K, so that $\sigma_{\mathrm{noise}} = \sqrt{k_B T_\text{noise} (f_s / 2)}$.

Using the results in \fref{fig:mb} to guide the conceptual design and construction of an ultimate CRES facility for measuring the absolute neutrino mass with T atoms, the general goals of the QTNM project are centered around: (1) the development of high phase-space density, low kinetic energy sources of T atoms and the implementation of methods to efficiently filter, confine and transport these atoms to the CRES measurement region; (2) the realisation a scalable approach to the construction of a CRES spectrometer that could, in the longer term, be implemented in a large scale facility to achieve a total instrumented volume of $\sim \qty{10}{\m\cubed}$; (3) exploiting atomic quantum sensors for high-precision minimally-invasive magnetic field mapping within the instrumented region of the CRES spectrometer; (4) the development of a multichannel antenna or a cavity based approach to collecting the cyclotron radiation with high efficiency, low background noise, and with controlled microwave systematics; and (5) the development of scalable multichannel quantum-noise-limited cryogenic microwave receivers that maximise the ratio of the cyclotron radiation signal from the $\beta$-decay-electrons to the background and amplifier noise, allowing precise microwave frequency metrology.

\begin{figure}[h]
\includegraphics[width=0.99\textwidth]{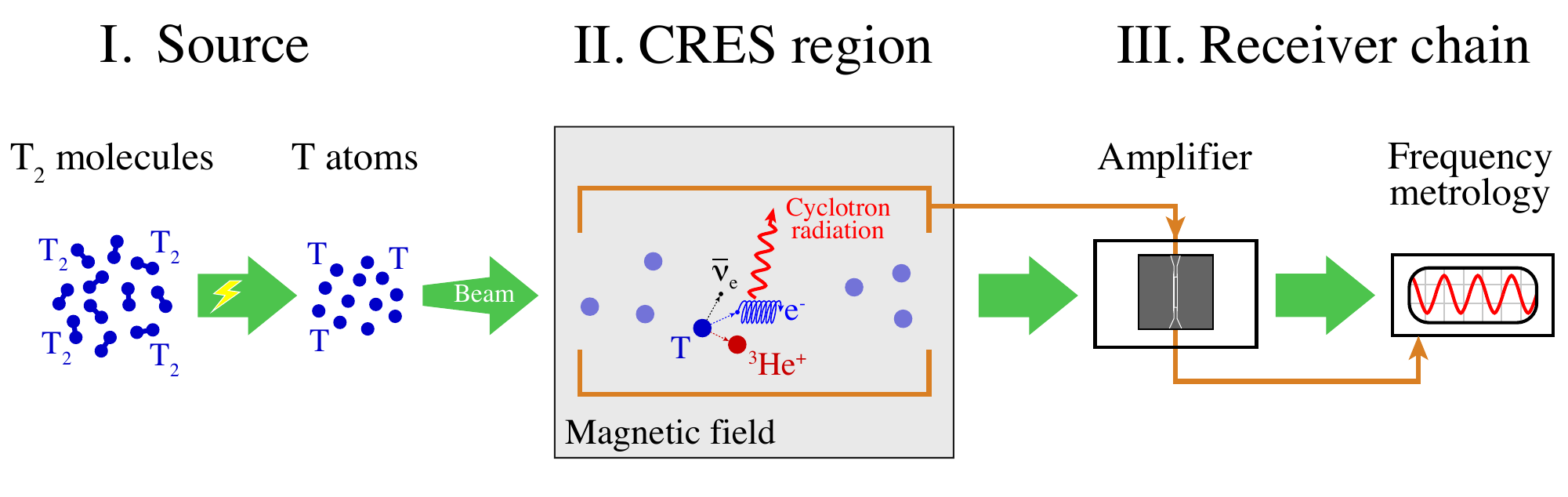}
\caption{
	Conceptual layout of the QTNM apparatus. 
	This comprises: (I) A high density source of T atoms produced by dissociation of T$_2$. 
	(II) A CRES region in which the cyclotron radiation from electrons generated by $\beta$-decay of T atoms in a homogeneous magnetic field is collected. The magnetic field is measured with atoms in superpositions of Rydberg states.
	(III) A receiver chain, containing quantum-noise-limited amplifiers, in which the cyclotron radiation will be amplified and measured.
}\label{fig:qtnmscheme}
\end{figure}

\subsection{Experimental approach}

% How QTNM will tackle the neutrino mass measurement
The general approach followed by the QTNM project is shown schematically in \fref{fig:qtnmscheme}. 
The main elements of this scheme are a high phase-space density source of T atoms (I) produced by the dissociation of T$_2$ molecules in an electric discharge. 
The T atoms are then transported to the CRES region (II) as a magnetically guided beam. 
In this cryogenically cooled electron spectrometer, a homogeneous background magnetic field is applied. 
This, and stray electric fields, are characterised by Rydberg-atom magnetometry and electrometry that can be implemented \emph{in situ} when the apparatus is operating at cryogenic temperatures using the T atoms themselves, or other species that cause minimal contamination. 
Electrons generated by T atoms that $\beta$-decay in the CRES region are detected by measuring the microwave radiation emitted as they undergo cyclotron motion in the magnetic field. 
The cyclotron radiation is collected by an ultra-low-noise receiver, which is mode matched as best as possible to the field distribution of the cyclotron radiation. 
In the receiver chain (III), the microwave radiation is amplified using purpose-developed quantum-noise-limited cryogenic amplifiers before down conversion to the radio-frequency region, and finally digitisation. 

Because of the large scale of the experimental apparatus that must ultimately be constructed, individual components are being developed, integrated, tested and refined sequentially. 
In the first phase of the project, high density cryogenic supersonic beam sources of H and D atoms are being developed and optimised. 
These sources are designed to be suitable for producing similar beams of T atoms. 
The motivation and considerations for choosing  these sources are discussed in \sref{sec:asource}. 
This phase of development also includes work on separating the ground-state atoms in the supersonic beams from the molecules, and implementing magnetic hexapole guides to transport them to the CRES region. 
Ultimately, additional magnetic guides and lenses will be used to transport the atoms between a series of spatially separated modular CRES regions in a scaled-up version of the apparatus. 
This could be achieved with long straight beam lines comprising arrays of coaxial CRES regions, or with CRES regions separated by curved guides in circular arrangements to allow the atoms to be recycled in a magnetic storage ring. As discussed in \sref{sec:confinement}, a prototype storage ring of this kind is being developed in the first phase of the project. 

To map magnetic and stray electric fields inside the CRES region with high absolute precision and with minimal disturbance to, or contamination of, the vacuum apparatus, atoms in high Rydberg states are used as quantum sensors. 
The techniques and protocols for this field mapping are also being developed and optimised in the first phase of the project, as discussed in \sref{sec:magnetometry}. 

In the CRES spectrometer, high-precision electron spectroscopy will be performed by measuring the frequency of the cyclotron radiation from $\beta$-decay electrons produced \emph{in situ}.
To maximise the cyclotron radiation signal, and therefore the precision with which the cyclotron frequency is determined, the electrons can be confined within the spectrometer which will be operated at a temperature of $\leq\qty{4}{K}$. 
This cryogenic environment also allows for the implementation of quantum-noise-limited superconducting amplifiers.
The spectrometer is described in further detail in \sref{sec:QTNMspec} and the development of quantum-limited superconducting amplifiers in \sref{sec:Qamps}.

The prototype QTNM CRES spectrometer module currently in development will form the core of the CRES demonstrator apparatus (CRESDA). 
This facility will be constructed to detect cyclotron radiation from single test electrons with energies up to $\sim20$~keV in an instrumented volume of $\mathcal{O}\left(\qty{10}{\cm\cubed}\right)$.
Initially, the microwave receiver within CRESDA will be constructed around cryogenic HEMT amplifiers, before moving to two-stage amplification with quantum-limited superconducting amplifiers.
CRESDA will allow tests to be performed to characterise and optimise receivers based on waveguides, cavities and antenna arrays with the goal of identifying the optimal configuration for use in a large-scale facility. 
Work will be carried out with magnetic fields in the CRES region ranging from \qtyrange{0.1}{1}{\tesla} (corresponding to cyclotron radiation frequencies from \qtyrange{2.7}{27}{\GHz}).
The single test electrons will be generated with high kinetic energies in their cyclotron degree of freedom by acceleration in a compact Penning trap. 
After acceleration, the electrons from this source will be magnetically transported to CRESDA. 

It is envisaged that the design, and many components, of CRESDA will be compatible with operation in an experiment with T atoms. 
This will allow a smooth evolution to a subsequent measurement phase, in which studies of the $\beta$-decay-electron spectrum of T will be carried out by CRES.
The initial development of CRESDA will take place in university laboratories, primarily in the UK. 
The development of each component will be informed by the expertise in handling and working with tritium at the Culham Centre for Fusion Energy (CCFE), which could act as a future host of a large-scale experiment. The ultimate large-scale CRES facility, with a working volume on the order of 10~m$^3$ which is required to be sensitive to values of $m_{\beta}$ below 100~meV/$c^2$, is expected to require international collaboration.

\section{Producing and detecting atomic hydrogen isotopes: H, D, and T}
\label{sec:asource}

To produce gases of H or D atoms for prototyping, or T atoms for an absolute neutrino mass measurement, it is necessary to dissociate H$_2$, D$_2$ or T$_2$. At threshold, this requires an energy of 4.46~eV~\cite{cheng18a}. Alternatively, a hydrogen bond in another stable molecule can be broken. The most widely used approaches to producing significant quantities of H and D atoms involve dissociating H$_2$ or D$_2$ by thermal cracking~\cite{olmstead23a}, or in an electric discharge~\cite{nagle47a,davis49a}, or by ultraviolet laser photolysis of NH$_3$ or ND$_3$~\cite{willitsch04a}. Because of the wider availability of T$_2$ than NT$_3$, high density sources of cold T atoms based on the dissociation of T$_2$ are expected to be most appropriate for next generation absolute neutrino mass experiments. 

\subsection{Available sources}

In the chemical and semiconductor industries, large quantities of H atoms are often produced by thermal cracking. This involves heating gaseous H$_2$ to $\sim2500$~K, i.e., thermal energies on the order of the threshold dissociation energy of the molecules~\cite{olmstead23a}. At these high temperatures dissociation occurs, and H atoms are produced with a high mean speed ($\langle v \rangle\simeq7\,000$~m/s), and broad speed distribution ($\sigma_v\simeq3\,000$~m/s). This approach to H and D atom production is robust and scalable, but the broad speed distributions make it challenging to control the motion of the atoms, or cool them to temperatures at which the typical kinetic energies are lower than required for a measurement of $m_{\beta}$ as outlined in \sref{sec:qtnm:goals}.

Continuous, effusive sources of large quantities of H atoms based on radio-frequency (RF) discharges of H$_2$ were successfully used in experiments that led to the observation of the first atomic hydrogen Bose-Einstein condensate~\cite{fried98a}, and are commonly used as sources for high-precision laser spectroscopic measurements, e.g., of the 1S--2S transition frequency~\cite{parthey11a}. In these experiments, cold atoms with speeds below $\sim100$~m/s, i.e., $E_{\mathrm{kin}}<0.05$~meV or $E_{\mathrm{kin}}/k_{\mathrm{B}}<1$~K, are generated by collisions with cryogenically cooled surfaces. 

Cooling H atoms by surface collisions is effective because the surface binding energy and residence time is generally sufficiently low to preclude significant losses by recombination to form H$_2$ before desorption. However, this cooling mechanism can be difficult to extend to produce cold, high phase-space density gases of D (or T) because the higher mass results in stronger surface binding, longer residence times and consequently greater losses by recombination~\cite{arai98a,mosk01a,steinberger04a}. To prepare cold gases of these heavier isotopes of hydrogen, it is therefore preferable to use alternative approaches. 

High phase-space density samples of H atoms can be generated in pulsed supersonic beams by electric discharges of H$_2$~\cite{scheidegger22a}. A schematic representation of such a source, in which a DC discharge is seeded with electrons from a heated tungsten filament, is shown in \fref{fig:HDTbeams}(a). This approach can be extended to generate beams of D or T atoms in discharges of D$_2$ or T$_2$. These supersonic beams typically have comparatively high mean speeds ($\langle v\rangle\sim1000$~m/s; $E_{\mathrm{kin}}\sim5$~meV), but narrow speed distributions ($\sigma_v \lesssim 50$~m/s). These characteristics mean that atoms in these beams are amenable to guiding, deceleration or trapping using inhomogeneous electric or magnetic fields~\cite{hogan11a}. In the QTNM project it is foreseen to develop high phase-space density supersonic sources of ground-state H, D and ultimately T atoms. The motion of these atoms will be controlled using inhomogeneous magnetic fields to confine samples for the measurement of $m_{\beta}$ by CRES. 

\subsection{Supersonic beams}

Supersonic beams are generated by the adiabatic expansion of a gas from a high pressure reservoir into vacuum through an aperture which is longer than the mean free path of the particles in the reservoir. In this situation, collisional cooling of the gas occurs as it exits the reservoir, and the thermal energy in the reservoir is converted into the directed motion of the particles in the beam. Supersonic beams can be operated continuously or in a pulsed mode with similar phase-space characteristics. The latter reduces the gas load on the vacuum system, and is therefore more appropriate for prototyping and system development. 

The mean longitudinal speed $\langle v\rangle$, longitudinal velocity spread, $\sigma_v$, and particle number density, $N$, in a supersonic beam depends on the temperature $T_0$, and pressure $P_0$ in the reservoir from which it emanates. They also depend on the diameter of the aperture through which the expansion occurs, and the propagation distance in the vacuum chamber~\cite{ashkenas66a,haberland85a}.

The particular gas species in a supersonic beam affects the phase-space properties through the individual particle mass, and the ratio, $\gamma$, of the heat capacity at constant pressure to that at constant volume. For beams containing H, D or T atoms produced by dissociation of H$_2$, D$_2$ or T$_2$, the supersonic expansion into the vacuum chamber is that of a gas of diatomic molecules ($\gamma = 7/5$) with mass $m_{\mathrm{H}_2} = 2.01588\,m_{\mathrm{u}}$, $m_{\mathrm{D}_2} = 4.02820\,m_{\mathrm{u}}$ or $m_{\mathrm{T}_2} = 6.03210\,m_{\mathrm{u}}$, respectively~\cite{nistweb23a}. This is because at the time of the initial expansion, only molecules are present. The H, D or T atoms subsequently produced after dissociation may therefore be considered to be seeded in the H$_2$, D$_2$ or T$_2$ beams.

\begin{figure}[t!]
\centering
\includegraphics[width=0.9\textwidth]{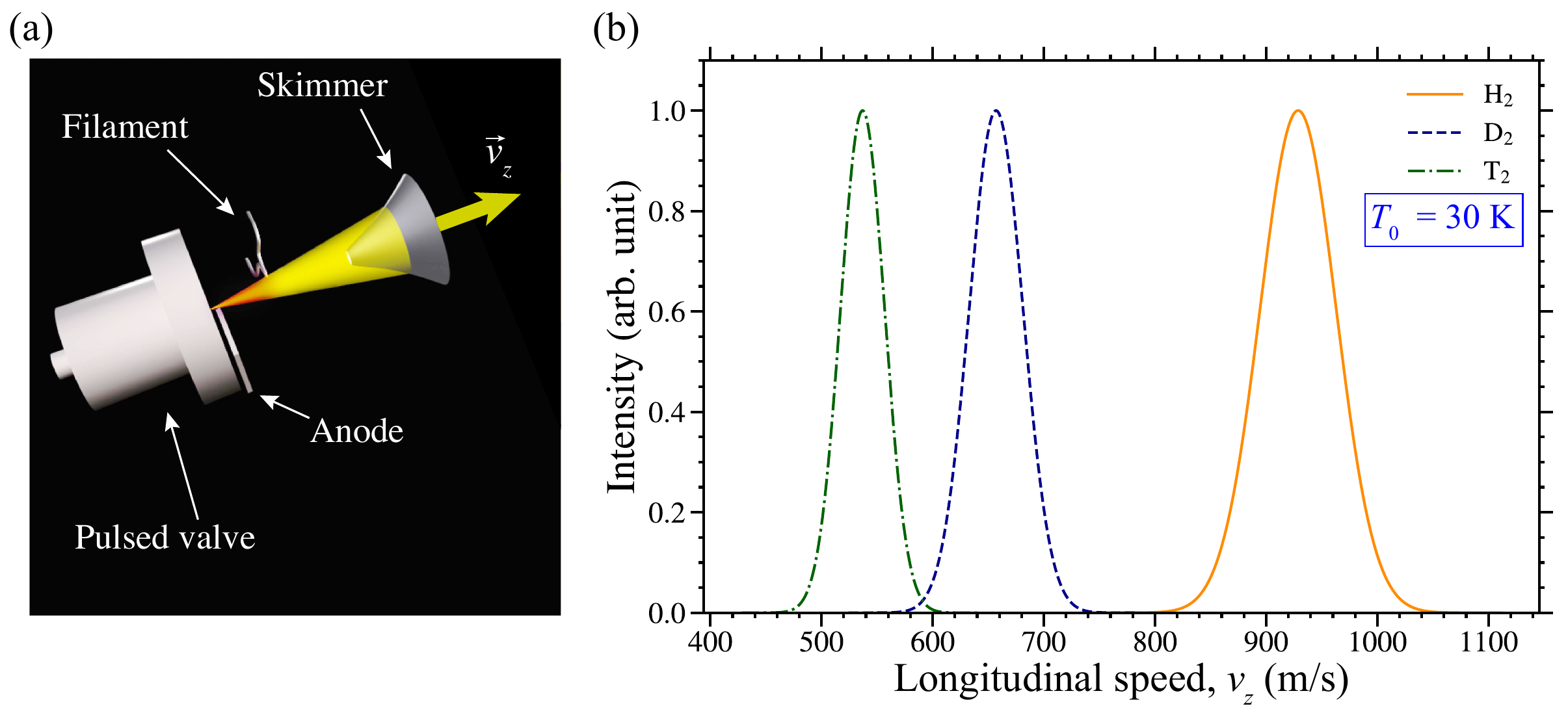}
\caption{(a) Schematic diagram of a pulsed supersonic beam source of H atoms (adapted from \cite{hogan12a}). (b) Calculated longitudinal speed distributions of supersonic beams of H$_2$, D$_2$ and T$_2$  emanating from a source operated at 30~K. H, D or T atoms with similar speed distributions can be generated in these beams by dissociation in an electric discharge.}\label{fig:HDTbeams}
\end{figure}
By way of example, in supersonic beams of H$_2$, D$_2$ or T$_2$ emanating from a nozzle maintained at $T_0=30$~K and $P_0 = 1$~bar, and with an orifice diameter of $200~\mu$m: the spread in kinetic energies in the moving frame of reference of the beams, at a distance of $10$~cm from the source, is $\Delta E_{\mathrm{kin}} \simeq m\,\sigma_z^2 / 2 \simeq 2~\mu$eV ($\Delta E_{\mathrm{kin}}/k_{\mathrm{B}}\simeq20$~mK); the mean longitudinal speeds are $\langle v\rangle = 910$, 660 and 540~m/s; and the standard deviations in these speeds are 18, 13 and 10~m/s in the cases of H$_2$, D$_2$ or T$_2$, respectively. 
The typical longitudinal speed distributions of these beams can be seen in \fref{fig:HDTbeams}(b). 
Under these conditions, the molecule number densities at a distance $10$~cm from the nozzle orifice are $N\sim10^{14}$~cm$^{-3}$. Dissociation of the molecules in these beams leads to the production of atoms with speed distributions that are similar to those of the molecular beam in which they are entrained. 
The number densities of atoms in the beams depend on the dissociation efficiency, and effects of recombination during propagation away from the source. 
Optimisation of these aspects of such a source of H, D and ultimately T atoms represents a major part of the QTNM project. 

The above parameters for the operation of supersonic hydrogen beams are, in general, compatible with the operating conditions and T$_2$ inventories currently available, for example, in the UK at CCFE, or in Germany at the Tritium Laboratory Karlsruhe. 
These research facilities maintain inventories of 10 -- 100~g of T$_2$, which is recovered and recycled after use (see, for example,~\cite{sturm21a}). 
We expect this quantity of T$_2$ to be sufficient for our planned experiments, including those that would reach the highest anticipated sensitivity in \fref{fig:mb}. 

\subsection{Detection of ground-state atoms}
\begin{figure}[t!]
\includegraphics[width=0.99\textwidth]{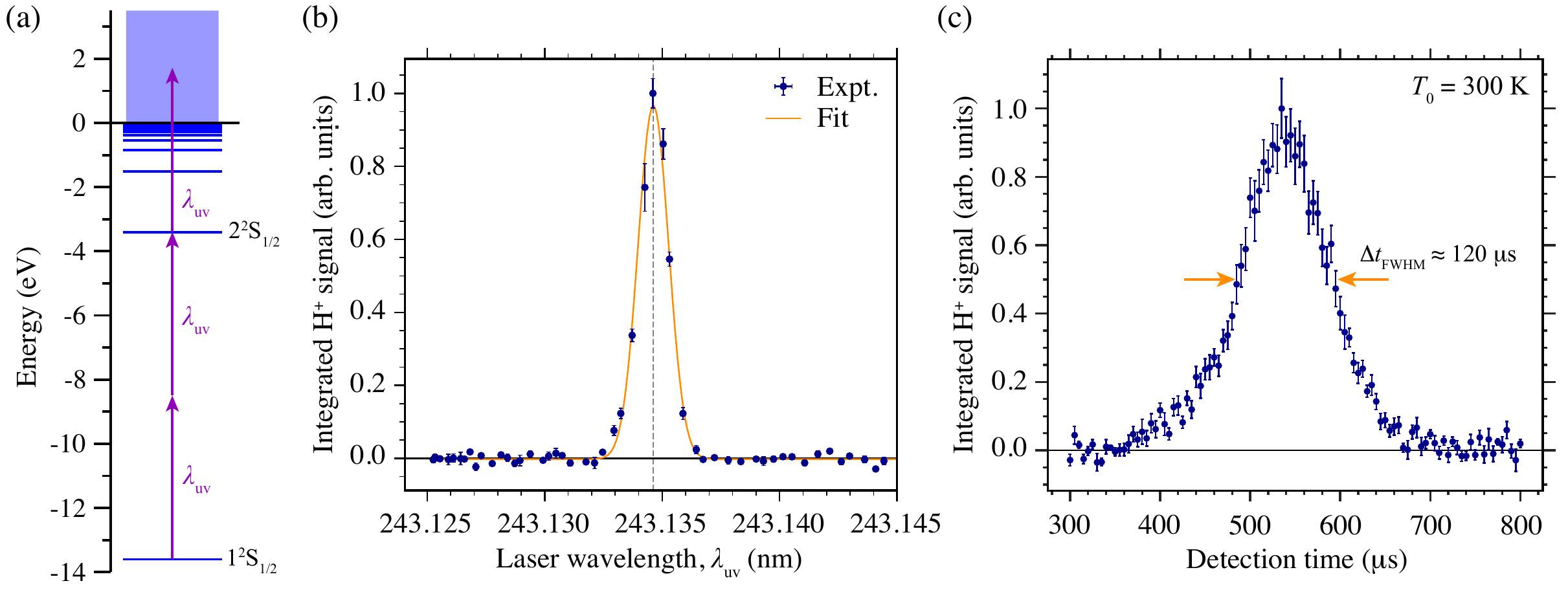}
\caption{(a) 2+1 REMPI scheme for detection of ground state atomic hydrogen isotopes. (b) REMPI spectrum of the 1S--2S transition in H recorded with laser radiation obtained by frequency doubling the output of a nanosecond Nd:YAG-pumped pulsed dye laser. (c) Time-of-flight distribution of ground-state H atoms generated by dc electric discharge in a pulsed supersonic beam of H$_2$, and detected by 2+1 REMPI at $\lambda_{\mathrm{uv}} = 243.1346$~nm.}\label{fig:REMPI}
\end{figure}
Detection of ground state H, D or T atoms, and \emph{in situ} determination of particle number densities and translational temperatures can be carried out efficiently and reliably by laser spectroscopic methods. For example, resonance-enhanced multiphoton ionization (REMPI) can be performed with a single laser tuned to resonance with the two-photon 1S--2S transition~\cite{bjorklund78a,downey89a}. This three photon ionization scheme is depicted in \fref{fig:REMPI}(a). The 1S--2S transition occurs at a wavelength $\lambda_{\mathrm{uc}} = 243.1346$, 243.0685 and 243.0465~nm in H, D, and T, respectively~\cite{parthey11a,parthey10a,nistHD05a}. In each of these cases, the subsequent absorption of a third photon at the same wavelength leads to ionization. By way of example, a 2+1 REMPI spectrum of the 1S--2S transition in atomic hydrogen recorded using radiation from a commercial nanosecond Nd:YAG-pumped, pulsed dye laser is presented in \fref{fig:REMPI}(b). When driven using light from these types of lasers, saturation of this REMPI process is typically approached with pulse energies of $\sim1$~mJ in beams focused to full-width-at-half-maximum waists of $\sim50~\mu$m. 

An example time-of-flight distribution of a pulsed supersonic beam of ground-state H atoms detected by 2+1 REMPI with collection of the resulting H$^+$ ions on a microchannel place detector is shown in \fref{fig:REMPI}(c). This beam was generated by dissociation of H$_2$ at the exit of a General Valve Series 99 pulsed valve (0.5-mm-diameter orifice, $P_0 = 3$~bar) operated at room temperature. This pulsed beam has a mean longitudinal speed close to 2800~m/s. The DC discharge used was similar to that depicted schematically in \fref{fig:HDTbeams}(a) with a sharp metal anode operated at a potential of $+200$~V which was positioned $\sim1$~mm from the front surface of the valve~\cite{halfmann00a}. A heated tungsten filament located $\sim25$~mm downstream from the valve was used as a source of electrons to initiate the discharge. As seen in \fref{fig:HDTbeams}, more slowly moving beams with narrower velocity distributions can be generated using similar electric discharge arrangements but a valve cooled to low temperatures.

\section{Controlling the motion of and confining atomic hydrogen isotopes}\label{sec:confinement}

For a measurement of the absolute neutrino mass with T atoms, it is desirable to confine the atoms within the experimental apparatus in a way that allows the detection of electrons generated by $\beta$-decay over long timescales. However, in contrast to experiments with T$_2$ -- which can be confined by collisions with the walls of a gas vessel because they are not radicals with unpaired electrons and therefore not reactive  -- confinement of ground-state T atoms is most effectively achieved using externally applied fields. 

The isotopes of atomic hydrogen, including T, are paramagnetic. Their non-zero ground-state magnetic dipole moment arises as a result of the presence of the single unpaired electron. Forces can therefore be exerted on them using inhomogeneous magnetic fields~\cite{hogan11a}. However, if these neutral atoms are excited to high Rydberg states, that possess large static electric dipole moments and linear Stark energy shifts, in externally applied electric fields, forces can also be exerted on them using inhomogeneous electric fields~\cite{hogan16a}. These features have previously been exploited to prepare cold magnetically trapped ground-state H and D atoms~\cite{hogan_2008b,wiederkehr_2010a}, and electrostatically trapped H and D atoms in high Rydberg states~\cite{hogan_2008a,hogan_2013a}. To guide, transport and confine the large quantities of T atoms in the QTNM project, it is currently expected to be most appropriate to work with atoms in their ground electronic state. With this in mind, the following discussion centres around the use of magnetic fields to control the motion of ground-state H and T atoms. 

\begin{figure}
\centering
\includegraphics[width=0.9\textwidth]{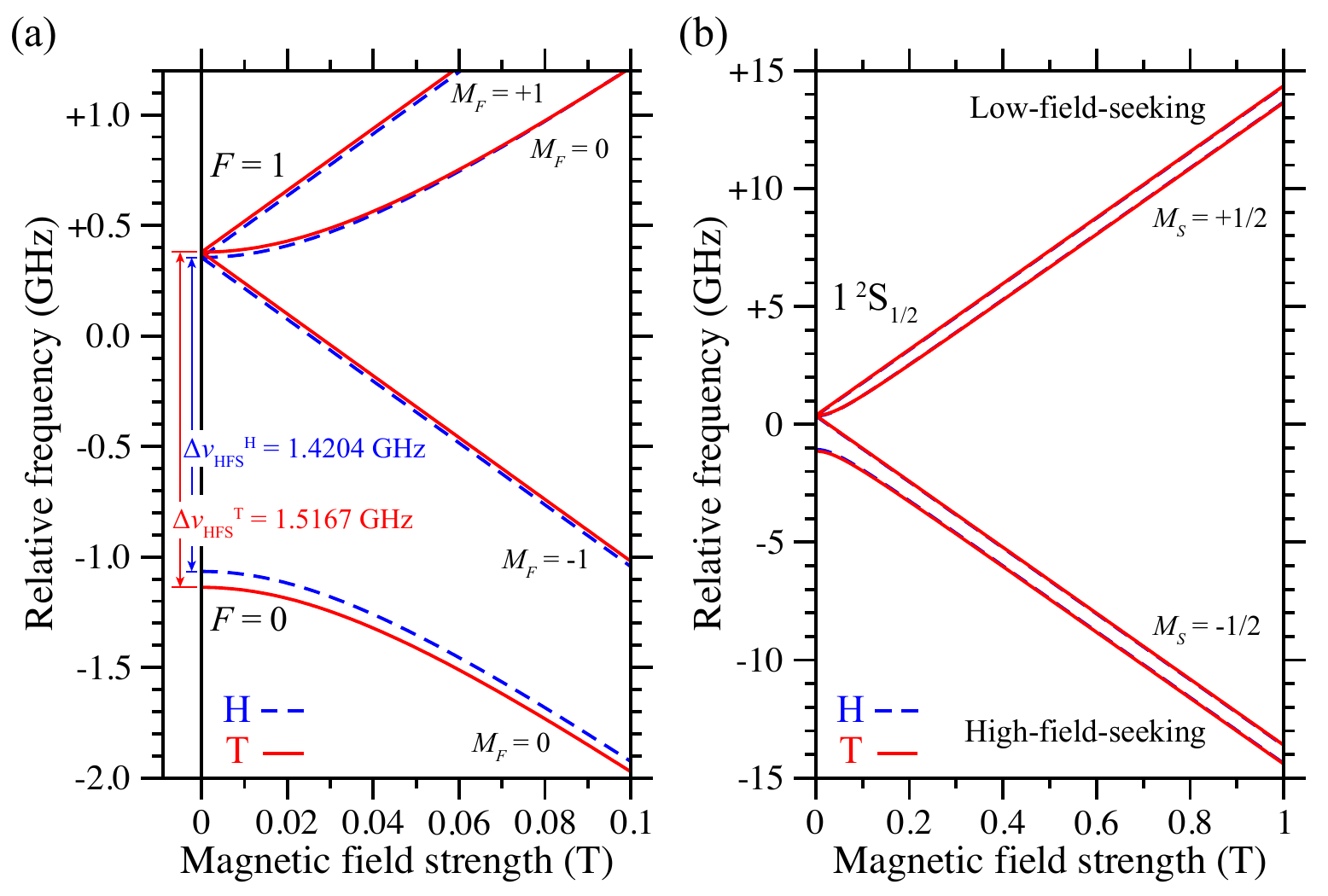}
\caption{Zeeman effect in the 1\,$^2\mathrm{S}_{1/2}$ ground state of H (dashed blue curves) and T (continuous red curves). (a) Hyperfine splitting and quadratic Zeeman shifts of $F=0$ and $F=1$ hyperfine levels in weak magnetic fields. (b) Linear Zeeman shifts of low- and high-field-seeking sublevels in strong fields.}\label{fig:Zeeman}
\end{figure}

% ----------------------------------------------------------------------
% ----------------------------------------------------------------------
% ----------------------------------------------------------------------

\subsection{Guiding, decelerating and trapping ground-state H and T atoms using inhomogeneous magnetic fields}

The effect of magnetic fields on the energy-level structure of the 1\,$^2\mathrm{S}_{1/2}$ ground levels of H and T are shown in \Fref{fig:Zeeman} (dashed blue and continuous red curves, respectively). In these figures, states with positive (negative) energy shifts are associated with situations in which the magnetic dipole moments, $\vec{\mu}_{\mathrm{mag}}$, of the atoms are oriented antiparallel (parallel) to the external magnetic field vectors. The linear Zeeman shifts seen in \fref{fig:Zeeman}(a) for the upper hyperfine sublevels with $F=1$ and $M_F=\pm1$ ($F$ is the total angular momentum quantum number including nuclear spin, and $M_F$ represents the projection of total angular momentum vector $\vec{F}$ onto the axis defined by the external magnetic field), and for the $M_F=0$ sublevels with $F=0$ and~1 in fields larger than $\sim0.05$~T, and which dominate in \fref{fig:Zeeman}(b), result from the magnetic dipole moment $\mu_{\mathrm{mag}}=\mu_{\mathrm{B}}$ associated with the spin of the unpaired electron ($\mu_{\mathrm{B}}$ is the Bohr magneton).

As seen in \fref{fig:Zeeman}, the energy-level structure and effects of magnetic fields on the 1\,$^2\mathrm{S}_{1/2}$ levels in H and T are, in general, similar. The proton and the triton (tritium nucleus) both have a nuclear spin quantum number of $I=1/2$. H and T therefore both possess ground-state hyperfine levels with total angular momentum quantum numbers including nuclear spin of $F=0$ and 1. The magnetic dipole moments of the proton, $\mu_p = 2.793\mu_{\mathrm{N}}$~\cite{schneider17a}, and triton $\mu_t = 1.067\,\mu_{\mathrm{p}}$~\cite{duffy59a} are similar ($\mu_{\mathrm{N}}$ is the nuclear magneton). Consequently, the hyperfine intervals of $\nu_{\mathrm{HFS}}^{\mathrm{H}} = 1.4204$~GHz and $\nu_{\mathrm{HFS}}^{\mathrm{T}} = 1.5167$~GHz between the $F=0$ and $F=1$ levels in H and T are also similar~\cite{anderson60a}. Together, these properties of the two hydrogen isotopes result in Zeeman shifts of the $M_{F} = 0$, $\pm1$ sublevels that differ slightly in weak magnetic fields [\fref{fig:Zeeman}(a)] but are almost indistinguishable on the scale shown in \fref{fig:Zeeman}(b) in stronger fields. It is worth noting that $I=1$ for the deuteron (deuterium nucleus), and the D atom has a significantly reduced ground-state hyperfine splitting compared to that in H and T. Consequently, over the range of magnetic fields in \fref{fig:Zeeman}(a) -- which are relevant for state selection and spin polarization -- the Zeeman effect in D is quite different. However, in stronger fields in which the electron magnetic moment dominants the Zeeman interaction, the energy level shifts are again comparable. 

The linear Zeeman shifts of energy levels in \fref{fig:Zeeman} in a magnetic field, $\vec{B}$,  can be expressed as $E_{\mathrm{Zeeman}} = -\vec{\mu}_{\mathrm{mag}}\cdot\vec{B}$. If this field is inhomogeneous, the gradient of the resulting Zeeman potential results in a force $\vec{f}_{\mathrm{Zeeman}} = -\nabla\,E_{\mathrm{Zeeman}} = \vec{\mu}_{\mathrm{mag}}\cdot\nabla\vec{B}$. This force leads to the acceleration of atoms in states with positive Zeeman shifts towards regions in space with lower field strengths. Atoms in these states are therefore often referred to as ‘low-field-seeking' (LFS). On the other hand, atoms in states with negative Zeeman shifts are accelerated toward regions of high field strength and are referred to as ‘high-field-seeking' (HFS). Atoms in LFS states, travelling in beams with narrow speed distributions, are well suited for guiding, deceleration and trapping using inhomogeneous magnetic fields~\cite{hogan11a}. This is because guides, decelerators and traps with large phase-space volumes over which particles undergo stable trajectories at constant phase space densities are more readily realised if the forces applied act toward regions in space of low field strength, rather than regions of high field close to saddle points in these field distributions.  

The largest quantity, and highest number density, gases of cold magnetically-trapped ground-state H atoms that have been reported were those achieved in experiments to produce Bose-Einstein condensates~\cite{fried98a,killian98a}. In that work, the requirement for cold, or slowly-moving atoms was set by the maximal achievable magnetic trap depths. For the typical Ioffe-Pritchard magnetic traps used, with a difference in magnetic field strength between their minima and saddle-points of $B_{\mathrm{trap}}\sim1$~T atoms in LFS Zeeman sublevels could be confined provided their kinetic energies were lower than $E_{\mathrm{Zeeman}} = \mu_{\mathrm{B}}\,B_{\mathrm{trap}} = 9.27\times10^{-24}~\mathrm{J}\equiv58~\mu\mathrm{eV}$. In this situation, $E_{\mathrm{Zeeman}}/k_{\mathrm{B}} = 0.67$~K and the corresponding maximal speed of atoms that could be trapped was $\sim100$~m/s. In those experiments, H atoms generated in an RF discharge of H$_2$ were cooled by collisions with cold surfaces. This allowed on the order of $\sim10^{11}$ LFS spin-polarized H atoms to be trapped at typical number densities of $10^{14}$~cm$^{-3}$ in each cycle of the experiment. However, as mentioned in \sref{sec:asource}, this experimental approach is challenging to extend to the preparation of cold magnetically trapped gases of D or T atoms.

The challenges in efficiently cooling T atoms for an absolute neutrino mass measurement can be avoided if translationally cold, supersonic beams are transported through the experimental apparatus in two-dimensional guides, or are decelerated and subsequently trapped in three-dimensional traps using inhomogeneous magnetic fields. Cold samples of magnetically trapped H and D atoms in LFS Zeeman sublevels have been prepared in this way using the methods of multistage Zeeman deceleration~\cite{vanhaecke07a,narevicius08a}. In this process, atoms moving at initial longitudinal speeds of $\sim500$~m/s in pulsed supersonic beams were decelerated using a series of pulsed magnetic field gradients generated in a set of co-axial solenoids (7~mm inside diameter, 64 windings of 300-$\mu$m-diameter copper wire, 11~mm long) by rapidly pulsing currents of up to 300~A. To decelerate the atoms to rest in the laboratory frame of reference and magnetically trap them, 12 solenoids were operated so that a reduction in kinetic energy of $\Delta E_{\mathrm{kin}}\simeq0.1$~meV was achieved upon passing through each~\cite{hogan_2008b}. The temperature of the gas of atoms that was trapped in this way was $\sim150$~mK and reflected the phase space acceptance of the decelerator. The same approach has been extended to use a set of 24 coaxial solenoids to decelerate and magnetically trap pulsed supersonic beams of ground state D atoms at a similar translational temperature~\cite{wiederkehr_2010a}. Since these techniques only rely on the forces exerted on the atoms in inhomogeneous magnetic fields and not collision processes for cooling, they could also be readily implemented to prepare cold trapped samples of ground state T atoms.

\begin{figure}
\centering
\includegraphics[width=0.85\textwidth]{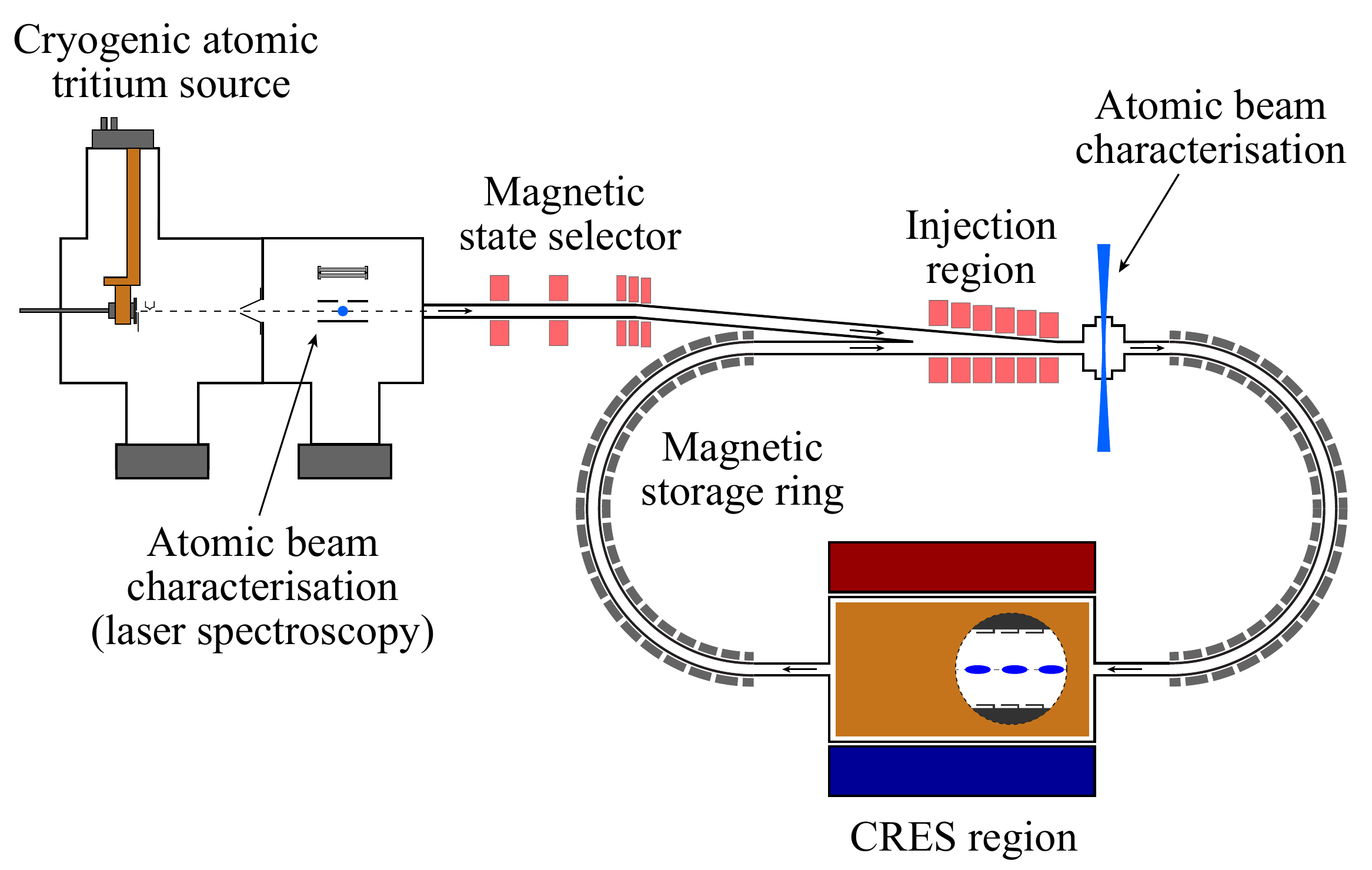}
\caption{Schematic diagram of the QTNM storage ring concept for a modular and scalable apparatus to measure the absolute neutrino mass by CRES.
The magnetic and electric fields within the CRES region are precisely mapped with atoms in superpositions of Rydberg states.}\label{fig:ring}
\end{figure}

\subsection{A modular magnetic storage ring for neutrino mass measurements}

The measurement of the $\beta$-decay electron spectrum of T atoms to a precision of $10 - 100$~meV by CRES, to determine the absolute neutrino mass can be achieved without preparing samples with temperatures below 1~K, i.e., $E_{\mathrm{kin}} \sim 100~\mu$eV. Therefore, three-dimensional trapping of the atoms is not essential. Such a measurement precision can be achieved with atoms moving slowly in beams with kinetic energies on the order of $E_{\mathrm{kin}} = 1 - 10$~meV. Taking this into consideration opens the opportunity to maximise the quantity and number density of atoms in the measurement region of the apparatus at any given time without the simultaneous need to prepare and maintain gases at low temperatures. In the QTNM project it is foreseen to achieve this in a modular and scalable CRES apparatus constructed in a long straight magnetically guided beam line, or a magnetostatic storage ring for LFS spin-polarized ground-state T atoms. A schematic diagram of the QTNM storage ring concept which is currently under development is shown in \fref{fig:ring}. 

% ---------------------------------------------
% ---------------------------------------------

In the atomic storage ring apparatus, T atoms pass through a magnetic state selector. This will allow them to be spatially separated from the molecules in the beams and spin polarized to minimise subsequent collision-induced losses from the ring~\cite{stoof88a}. The resulting atoms will be injected into the storage ring using an arrangement of permanent magnets to generate an inhomogeneous magnetic field distribution that acts as a beam combiner for atoms in LFS states. A prototype version of this storage ring is currently under development. This ring comprises two 180$^{\circ}$ curved sections, each with a radius of curvature of 0.6~m. These each contain 60 permanent magnet guide segments in which hexapole magnetic field distributions are generated by sets of 12 magnets in Halbach arrays. These guide segments are each 5~cm long and have an open bore with an inner diameter of 5~mm through which the atoms propagate. The injection region in this prototype is located in a straight part of the ring between the 180$^{\circ}$ curved sections. It is ultimately foreseen to install a modular CRES detection region, comprising superconducting solenoid magnets to generate a strong homogeneous magnetic field and microwave receivers, in the straight section on the opposite side of the ring. Atom optics elements (not shown) including, for example, magnetic lenses and guides are required to transport and collimate the atomic beam between the curved sections of the ring to the CRES region. Because the Zeeman effect in the ground states of H and T is similar (see \fref{fig:Zeeman}), the state-selector and storage ring will operate in a similar way and with a similar efficiency for these two isotopes of atomic hydrogen. It is therefore planned to first develop and characterise the operation of the prototype ring primarily with H atoms moving at mean longitudinal speeds of $\sim900$~m/s ($E_{\mathrm{kin}}\simeq4$~meV) before ultimately moving to work with T atoms emanating from the same source at mean longitudinal speeds of $\sim500$~m/s ($E_{\mathrm{kin}}\simeq4$~meV). 

In this storage ring concept, the T atoms that act as the source of $\beta$-decay electrons are those located at any given time within the CRES region. The typical number of atoms contributing to the neutrino mass measurement therefore corresponds to the time-averaged number of atoms in this instrumented region. During the initial development of the apparatus, the H or T atom source will be operated in a pulsed mode. However, in the longer term, it is foreseen to generate continuous supersonic beams of T atoms in the same type of source to maximise the time-averaged number of atoms in the CRES region. To scale the apparatus to allow the $\beta$-decay spectrum of larger quantities of T atoms to be recorded, and increase the count rate close to the end point, the separation between the 180$^{\circ}$ curved sections of the ring can be increased so that multiple equivalent CRES modules can be installed in the straight sections on each side.

To achieve the instrumented volumes of $10^{-3}$, 0.05 and 10~m$^{3}$ considered in the sensitivity analysis in \fref{fig:mb}, it will be necessary to exploit the full transverse spatial distribution of atoms generated in the source. For a beam that transversely fills a 0.5-m-long CRES region with an inner diameter of 5~cm, an instrumented volume of $10^{-3}$~m$^3$ is expected to be achievable in a single CRES module. To move to larger scale versions of the apparatus, it will be necessary to increase the individual CRES module size to an inner diameter of $\sim10$~cm and a length approaching 1~m. This would allow a total instrumented volume of 0.05~m$^{-3}$ to be achieved in a ring, or beam line, comprising $\sim6$ CRES modules. A further increase in the length of the individual CRES regions by a factor of 2 while maintaining an inner diameter of $\sim10$~cm (or an increase in both the length and inner diameter by factors of 2) would allow total instrumented volumes on the order of 10~m$^{3}$ to be achieved with 640 (160) modules which could be arranged in multiple parallel beam lines, or storage rings.

% ------------

\section{Atomic magnetometry and electrometry}
\label{sec:magnetometry}

The energy of an electron liberated by the radioactive decay of a parent atom is best measured through CRES, where a strong (0.1 - 1.0 T, depending on frequency) static magnetic field is maintained to force the the electron into a circular, or at least helical, orbit. 
However, to achieve the spectroscopic precision required, it is essential that the magnetic field is uniform, known, and continuously measured. 
A challenge with the static magnetic field, is that it must be known throughout the whole of the working volume; unless extreme spatial stability can be ensured.
An additional problem is that even low-level stray electric fields can change the dynamics of the electron, so these must be kept to a minimum. 
The electric fields in the spectrometer can in principle be set to zero by suitable shielding, but stray fields caused by patch potentials, adsorbates and imperfections on the cryogenically cooled surfaces must also be accounted for to minimise systematic errors. 
Ideally, these electric fields must also be characterised, if only on a statistical basis.

\begin{figure}
\includegraphics[width=0.99\textwidth]{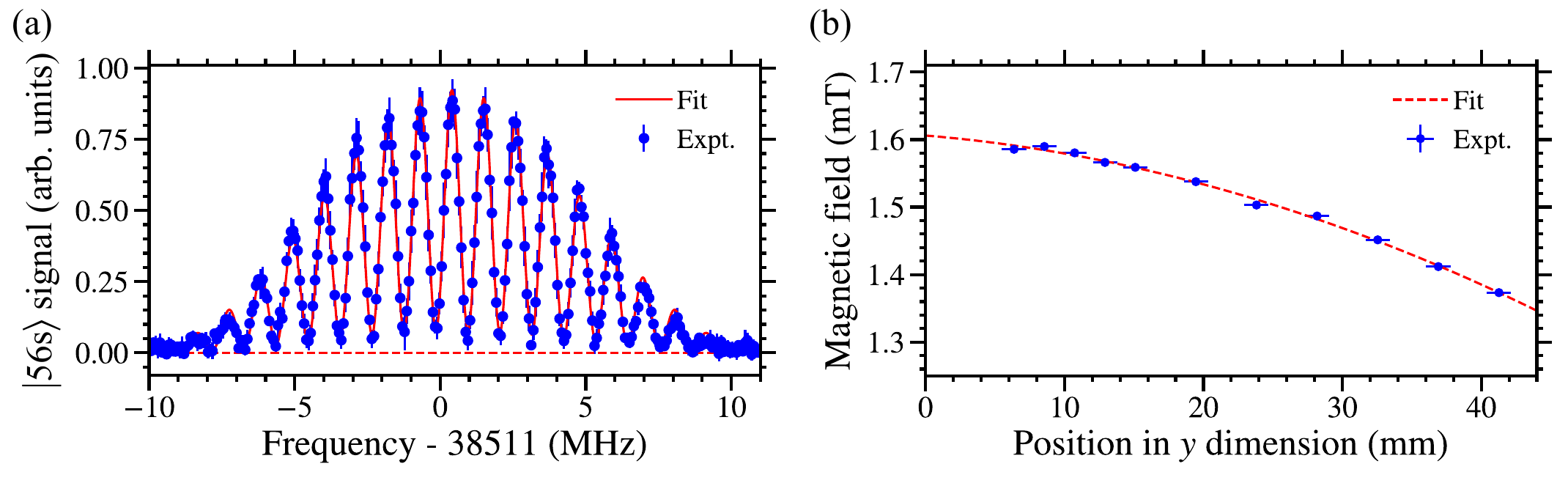}
\caption{(a) Ramsey spectrum of the transition between the $n=55$ and $n=56$ circular Rydberg states in helium used for absolute static field magnetometry. (b) Example of a one-dimensional magnetic field map measured using atoms in circular Rydberg states as quantum sensors. Adapted from~\cite{zou23a}.}\label{fig:mag}
\end{figure}

The measurement of static magnetic fields to an absolute precision of $\sim1~\mu$T, and static stray static electric fields to an absolute precision of $\sim100~\mu$V/cm can be achieved over comparatively large regions ($0.1 - 1$~m) in an atomic tritium CRES apparatus by using the T atoms themselves, or other inert atoms such as helium which will not contaminate the measurement region in a detrimental way, as quantum sensors. 
This magnetic and electric field sensing is achieved most effectively if the atoms are prepared in circular Rydberg states. 
These are highly-excited electronic states in which the excited electron has the largest possible azimuthal and orbital angular momentum quantum numbers, $m_{\ell}$ and $\ell$, for any given value of the principal quantum number $n$, i.e., $|m_{\ell}| = \ell = n-1$. 
This is because pairs of circular Rydberg states that differ in their values of $n$ by $\pm1$ represent quasi two-level systems with long coherence times, which allow for maximal measurement sensitivity. 

Circular Rydberg states in H, T or other suitable species, such as helium (He), can be prepared by laser photoexcitation in crossed electric and magnetic fields~\cite{delande88a,zhelyazkova16a}, or by multiphoton radio-frequency transfer from low-angular momentum Rydberg states after laser photoexcitation in zero field~\cite{hulet83a}. 
These states offer the advantage for absolute static field magnetometry and electrometry that their Zeeman and Stark energy shifts can be calculated analytically to the high precision required in a CRES-based measurement of the absolute neutrino mass. 
These energy-level shifts can be precisely measured by coherent Ramsey spectroscopy at microwave frequencies~\cite{ramsey50a}, with detection by quantum-state-selective electric field ionisation. 
By way of example, a Ramsey spectrum of the transition between the $n=55$ ($|55\mathrm{c}\rangle$) and $n=56$ ($|56\mathrm{c}\rangle$) circular Rydberg states in He is shown in \fref{fig:mag}(a). 
This spectroscopy technique involves applying a short pulse of microwave radiation at a frequency close to resonance with the circular-to-circular state transition frequency to first prepare the atom in a superposition of these states. 
This superposition state is then allowed to freely evolve and accumulate a phase that depends on the frequency interval, including effects of Zeeman or Stark shifts, between two Rydberg states. 
Finally, a second pulse of microwave radiation is applied to project the resulting superposition state on to the two basis Rydberg states. 
The interference fringes observed in the spectrum in \fref{fig:mag}(a) arise as a result of the difference in phase accumulated by the atomic superposition state and the microwave field in the free evolution time between these two pulses.

From a spectrum of the kind in \fref{fig:mag}(a), the resonance frequency can be determined from a least-squares-fit of a Ramsey spectral line shape function to the experimental data. 
This frequency can then be used in combination with measurements of additional transitions to different final Rydberg states to determine the vector components of the residual uncancelled stray electric fields at the location of the atoms to a precision of $\pm100~\mu$V/cm, and the magnetic field strength to an absolute precision of $\pm1~\mu$T~\cite{zou23a}. 

This approach to Rydberg atom magnetometry and electrometry can be performed in a pulsed mode with bunches of excited atoms with sizes of $\sim100~\mu$m, $2~$mm and $\sim100~\mu$m in the $x$, $y$ and $z$ dimensions, respectively. 
In pulsed supersonic beams of He emanating from a source operated at room temperature, these atoms travel through the magnetic field to be characterised at a speed of 2~mm/$\mu$s. 
Consequently, by making measurements at a range of time delays after Rydberg state photoexcitation, magnetic and electric fields at a range of positions along the axis of propagation of the atoms can be mapped. 
The result of such a one-dimensional magnetic field mapping procedure is shown in \fref{fig:mag}(b) where the measured fields over a distance of 40~mm are displayed with an absolute precision of $\pm1~\mu$T and a spatial resolution of $\pm0.5$~mm.

Work is underway in the current phase of the QTNM project to implement these methods to precisely characterise magnetic fields of up to $\sim0.5$~T, and the individual vector components of the stray electric fields under the conditions encountered in a CRES apparatus. 
Based on the results of the measurements in \fref{fig:mag}(b), it is anticipated that this approach to minimally-invasive atomic magnetometry and electrometry in a CRES spectrometer operated at temperatures below 4~K, can be extended with comparable absolute precision and spatial resolution to large CRES modules with dimensions on the scale discussed in \sref{sec:confinement}.

\section{The QTNM electron spectrometer}
\label{sec:QTNMspec}

The QTNM electron spectrometer is being designed to meet several requirements. It needs to: (a) detect tritium $\beta$-decay electrons in the energy range of interest close to 18.6~keV, (b) measure the frequency of the cyclotron radiation emitted by each of these electrons, and (c) provide sufficient information on the trajectories of these electrons that their kinetic energy can be obtained from their cyclotron frequency with a precision and accuracy suitable for determining the value of $m_\beta$.

To reach the required sensitivity to $m_\beta$ it is necessary to build a spectrometer with which the kinetic energies of 18.6~keV electrons can be determined to a precision of $\sim$0.1~eV. 
As discussed in \sref{sec:QTNM}, this corresponds to a precision of $2\times10^{-7}$ in the electron cyclotron frequency in a CRES spectrometer. 
Consequently, as indicated by \eref{eq:fVar}, the spectrometer must capture as much of the cyclotron radiation generated by each individual electron as possible. 
At present, the QTNM collaboration is exploring several approaches to collecting this microwave radiation, including arrays of antennas and resonant cavities. 
These different approaches have their own advantages and disadvantages, and the trade-offs are delicate.

The fundamental problem is that we must measure the energies of individual electrons, but the observation times are short due to trapping and scattering considerations, and so it is not possible to integrate the signal for extended periods to achieve high signal-to-noise ratios. 
To maximise the signal-to-noise ratio, the radiation collecting system must couple efficiently to each electron, regardless of where it appears in the field of view.
However, as discussed in~\cite{withington2024} there is a fundamental information-theoretic trade-off between coupling efficiency and field of view. 
The experimenter is therefore forced into using multiple receivers, which means multiple antennas or cavity probes, and these necessarily interact. Nevertheless, digitally cross correlating, in addition to autocorrelating, the outputs of an array of receivers, all of which sample the same volume, gives some spatial information in addition to spectral information. 
A major advantage of cross-correlating the outputs of a number of receivers is that the effects of mutually uncorrelated noise is reduced.
To minimise the system noise, ultra-low-noise cryogenically cooled amplifiers must be used, ideally quantum noise temperature limited, and the effective radiation temperature of any stray thermal background must also be minimised. 

Although outward-looking phased arrays have been developed extensively, inward-looking phased arrays are poorly studied. 
For example, the receivers interact strongly, leading to standing waves within the field of view, and therefore potential spatial sensitivity variations over short scale lengths. 
These also complicate the analysis of cross-correlated data. 
Additionally, the input of one receiver can see the noise leaving the inputs of the others, which can be higher than noise temperatures of the amplifiers themselves. 
A further key consideration is that the ports that allow the T atoms to enter and leave the spectrometer become a source of thermal background noise, and the receivers should, ideally, not couple to this background.

Arrays of antennas, and to a lesser extent multiprobe cavities, offer the attractive ability to spatially reconstruct the trajectories of electrons, allowing for the mitigation of known spatial effects, such as variations in the magnetic field.
However, as \cite{withington2024} discusses, for volumes of more than a few cubic wavelengths, a large number of array elements are required, increasing the total number of channels that must be instrumented. 

An alternative approach is to use resonant cavities that provide, in principle, the ability to collect a large fraction of the radiated electron power with potentially only one or two probes.
However, this does not overcome the fundamental information-theoretic limit: a large-volume cavity will have a large spectral density of resonant modes which couple to the cyclotron frequencies of interest unless there is a corresponding reduction in the magnetic field used.
This brings difficulties both from the decreased radiated power and from the challenges in maintaining equivalent magnetic field uniformity.
Cavities may provide the attractive prospect of increasing the radiated power above that of the free-space Larmor power, via the Purcell effect~\cite{Purcell1946}.
This effect has previously been utilised for the suppression and enhancement of cyclotron radiation in a cavity~\cite{gabrielse1985}.
A theoretical analysis of multi-probe cavities, which compliments the work done on free-space antennas~\cite{withington2024},  will be reported by QTNM shortly. 

To observe the cyclotron radiation from the $\beta$-decay electrons over sufficiently long timescales, it is expected to be necessary to confine them within the CRES spectrometer.
The problem is that even a mildly relativistic electron moves a long distance in the observation time required, unless it is injected on a circular rather than helical trajectory.
A long detection region can be envisioned, but ultimately there is a limit because of the trade-off between volume and coupling efficiency. 
However, the method used to achieve electron trapping must cause minimal perturbations to the electron's kinetic energy over the observation time.

One possible approach to confine the $\beta$-decay electrons is to use a pure magnetic trap. 
This could be implemented by creating a local minimum in the magnitude of a background magnetic field. 
A range of options have been modelled, and arrangements of solenoids that can produce a longitudinal trapping potential that does not introduce transverse inhomogeneities have been identified.
Additionally, stray electric fields within the spectrometer must also be controlled so that they do not lead to unwanted changes in the electron's kinetic energy.

\begin{figure}[ht!]
    \centering
    \includegraphics[width=\linewidth]{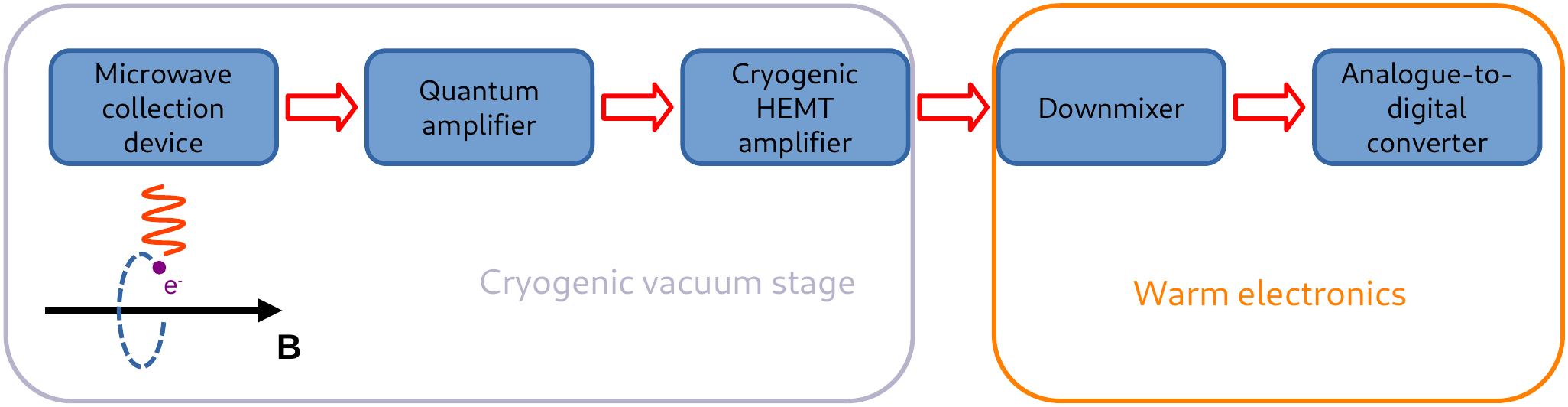}
    \caption{
    	Schematic diagram of the path taken by a CRES signal, indicating those parts of the signal processing chain that are operated at cryogenic temperatures and at room temperature.
    }
    \label{fig:signalFlow}
\end{figure}

\Fref{fig:signalFlow} shows a simple diagram of the path of a CRES signal in the apparatus.
Initially, an electron induces a signal in one or more microwave collection devices.
This signal is then passed to quantum-noise-limited amplifiers (see \sref{sec:Qamps}).
The performance of these devices approaches the quantum-limit where a half a quantum of noise is added per frequency cycle.
Following this, the signal is passed into a cryogenic HEMT amplifier~\cite{HEMTref}.
This two-stage amplification allows a high signal-to-noise ratio (SNR) to be maintained before the signal reaches the room-temperature electronics.

Previous experiments have reported detecting CRES events with system noise temperatures as high as \qty{145}{\K}~\cite{Project8_2015}. 
However, the QTNM collaboration aims to operate a spectrometer with a significantly lower noise temperature than this, resulting in a higher SNR. 
This will be achieved by operating the CRES region at a thermal temperature of \qty{4}{\K} and using quantum-noise-limited amplifiers for the first-stage of amplification at possibly \qty{0.1}{\K}. 
The noise temperature of importance, however, is the system noise temperature including any thermal background from the potentially warm environment, and there is no point in installing ultra-low-noise amplifiers if the radiometric background is high.

\Eref{eq:fVar} shows that
\begin{equation}
	\textrm{std}\left(\hat{\nu}_0\right) \propto \frac{1}{\sqrt{\text{SNR}}} \, .
\end{equation} 
Thus any reduction in the system noise temperature will result in an improvement in the achievable frequency precision for a given observation time, thus increasing the number of decay electrons which can be measured to a given precision and improving the experimental efficiency.

Additionally, the signal structure of a typical CRES event is expected to be more complicated than the chirped sine wave discussed in \sref{sec:qtnm:goals}.
A trapped electron with a pitch angle (the angle between the electron's momentum vector and the magnetic field) below $90^{\circ}$ will undergo periodic motion in the trap.
This motion may lead to a range of different magnetic fields being experienced, resulting in the emission of cyclotron radiation at a frequency given by the mean magnetic field experienced by the electron.
This motion can also result in frequency (from the Doppler effect) and amplitude (from the reception pattern of the microwave collection system) modulation of the signal, depending upon the microwave collection technology used.
The signal structure of CRES events is expanded upon further in~\cite{esfahani_2019}.
This modulation causes some power to be radiated in `sidebands' rather than at the main cyclotron frequency.
The position and strength of these `sidebands' relative to the main frequency can be used to deduce the motion of the electron in a trap and so calculate the energy of the electron from the detected radiation.
Therefore, ensuring that noise levels are low enough that sidebands are observable is a key goal for QTNM.

% Warm readout
The cyclotron frequencies expected in a typical CRES experiment exceed those which can be economically digitised for many channels.
Therefore, the experimental signals will be down-converted to the sub-\unit{\GHz} range.
This is possible without significant loss of information because most of the signal power is emitted at frequencies close to the cyclotron frequency. 
Digitisation will then be achieved using an analogue-to-digital converter with a sampling rate of $\sim\qty{1}{\GHz}$.

% Outline of triggering
The production of electrons from tritium $\beta$-decay occurs randomly, prompting an event-based approach to data collection for the experiment.
Since electrons in the region of interest close to the endpoint energy of the decay are very rare, a hardware- or software-based trigger will be needed to avoid collecting and processing large amounts of data containing no electrons of interest. A trigger for QTNM therefore needs to quickly identify the presence of these electrons in the background noise. Two trigger options are currently being explored for QTNM. These are a lock-in amplifier~\cite{Meade_1983} based approach and a matched filter~\cite{Turin1960} based approach.
These techniques are capable of detecting the presence of known signals in high noise environments, so could form the basis for a trigger system for CRES detection.
Both the matched filter and lock-in amplifier processes have previously been implemented on Field Programmable Gated Arrays (FPGAs)~\cite{Baker_2007, Stimpson_2019} and this implementation is also being tested by QTNM.

% Reconstruction
Having triggered on a CRES event, the next challenge is to determine the frequency spectrum of the radiation emitted and therefore the corresponding electron kinetic energy.
One complication is that it is necessary to know the frequency at the beginning of the chirp when it is assumed that the electron was first produced. 
Therefore, the error on the derived initial energy is sensitive to timing errors.
Furthermore, as mentioned above, the mean magnetic field experienced by each electron must be reconstructed.
Given the complex signal structure expected multiple advanced methods are being explored. These include a matched filter with a large template bank spanning the full range of electron parameters or machine learning techniques.

\section{Development of quantum-noise-temperature limited amplifiers}

\label{sec:Qamps}

QTNM's current strategy is to use either free-space microwave coupling or high-Q cavity coupling to the electrons from the $\beta$-decay of atomic tritium. The collaboration has investigated detection methods for the electron's cyclotron radiation across a range of magnetic fields from 0.5 to 1 T. The optimal frequency is a delicate trade-off between many competing considerations. To ensure a balance between  a high signal-to-noise ratio, a reduced level of systematic uncertainties (for example electromagnetic interference), and the availability of a wide range of commercial microwave components, the current focus is on a magnetic field of $\sim 0.7\,\mathrm{T}$ resulting in a primary observing frequency of \qty{18}{\GHz}.

An important consequence of this measurement technique is the need for a multi-channel readout system featuring a sizeable array of dozens, potentially hundreds, of ultra-low noise microwave receivers~\cite{withington2024}.
To reach sensitivities to the absolute neutrino mass below 0.1~eV/$c^2$ shown in \fref{fig:mb}, a CRES instrument with a large fiducial volume and a system noise temperature below \qty{2}{\K} is required, which is only possible if even lower noise temperature amplifiers are available. 
Near quantum-noise-limited amplifiers and larger arrays would be beneficial and are technologically achievable.

This level of performance is unlikely to be possible using conventional cryogenic amplifiers such as HEMTs.
Arrays with in excess of hundreds of channels with HEMT amplifiers are impractical, limiting the potential of future expansion to larger volumes and new measurement techniques.
Consequently, a central focus of QTNM activities has been to develop two types of slow-dissipation, arrayable superconducting amplifiers operating near the Standard Quantum Limit (SQL)~\cite{Caves_1982_quantum_limit,Clerk_2010_quantum_limit,Zhao_2021}. 
These are Superconducting Parametric Kinetic Inductance Amplifiers and Superconducting Low-inductance Undulatory Galvanometers (SLUGs). 
QTNM's goal is to determine the most suitable readout technology for a large volume CRES-based neutrino mass experiment.  

\subsection{Superconducting Parametric Amplifiers}

Superconducting parametric amplifiers come in a variety of forms: schemes based on the nonlinear kinetic inductance of thin-films; and schemes based on the nonlinear inductance  of superconducting tunnel junctions. 
Moreover, each of these categories subdivides into those configured as resonators, and those configured as nonlinear transmission lines. QTNM is concentrating on resonator kinetic inductance amplifiers because of their ease of mass production, satisfying the need for large arrays.

In general terms, superconducting kinetic inductance parametric amplifiers have demonstrated high gain $10-30 \,\mathrm{dB}$ whilst achieving added-noise close the SQL \cite{Eom_2012, Vissers_2016, Malnou_2020}. More recently, these amplifiers have also been used to achieve squeezing amplification and the generation of squeezed vacuum states \cite{Parker_squeezing_2022}. These remarkable gain and noise characteristics allow kinetic inductance parametric amplifiers to improve the performance of various quantum information and detector systems \cite{Ranzani_2018,Zobrist_2019,Vissers_2020}.

The basic mechanism is that, in the presence of a strong pump tone, nonlinearity in the inductance results in wave-mixing processes, which transfer energy from the strong pump tone to the weak signal tone, thereby achieving amplification \cite{zhao2022physics}. Superconducting parametric amplifiers in both travelling-wave and resonator geometries have been realised, but the resonator schemes have many advantages for narrow-band applications:  resonator amplifiers have lower power requirements on the pump tone and so are less prone to heating and microbridge hot-spot switching, they therefore should be less noisy, and above all they are easy to manufacture in bulk because they are less prone to lithographic defects such as shorts or breaks, which are notoriously challenging on the long transmission lines of travelling-wave structures \cite{Eom_2012,Shan_2016,zhao2022physics}. 

The nonlinear inductance is necessarily associated with a nonlinear resistance \cite{Hutchings1992_KK_relations,Zhao_2022_mechanisms_TWPA,Zhao_2023}, but the theory of resonator devices has been developed to a high level of understanding \cite{Thomas_2020,Thomas_2022}, which allows for the design and operation of resonators as amplifiers routinely, despite the fact that the frequency response of resonators displays complex hysteretic behaviour.

\begin{figure}[t!]
\includegraphics[width=1.0\linewidth]{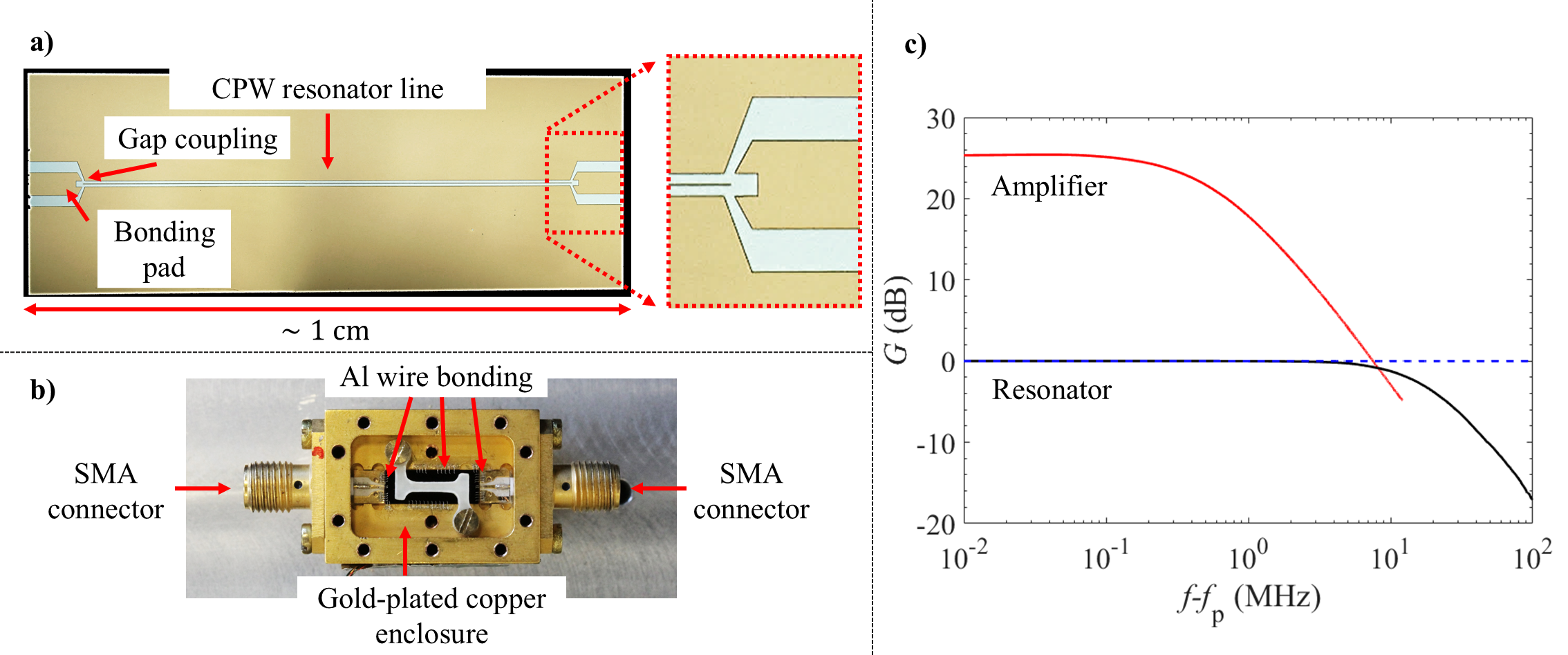}
\caption{\label{fig:Paramp} a) An example of NbN half-wave resonators fabricated for the QTNM project. The resonator comprises a length of coplanar waveguide capacitively coupled in series between two contact pads. b) Gold-plated copper enclosure with sub miniature version A (SMA) connectors for microwave measurement. c) Measurement of power gain $G$ as a function of difference frequency $f-f_\mathrm{p}$, where $f$ is the signal frequency and $f_\mathrm{p}$ is the pump frequency. The black line shows the transmission when there is no pump tone, i.e., the device is acting as a simple resonator; the red line shows the gain in the presence of a strong pump, i.e., the device is acting as a resonator amplifier.}
\end{figure}

Superconducting NbN thin-films are used for the amplifier material, using a fabrication process that optimises the relevant film properties. 
A comprehensive theoretical analysis on the operation of these resonator amplifiers has been conducted~\cite{Thomas_2022}. Guided by our analysis, many half-wave and quarter-wave resonator amplifiers have been successfully designed, fabricated, and operated in both transmission and reflection modes. These configurations achieve high, stable gain over several MHz of bandwidth \cite{Zhao_2023}. \Fref{fig:Paramp} shows an example of a NbN half-wave resonator, the resonator enclosure, and the measured power gain. As shown, for this particular resonator, high gain of $G >20\,\mathrm{dB}$ can be achieved over $\sim1\,\mathrm{MHz}$ of bandwidth. Larger bandwidth designs are now being studied, with \qty{5}{\MHz} being entirely feasible. 

Experimental results show that the reflection-mode design is capable of intrinsic pump removal \cite{zhao2024_intrinsic_separation}; additionally, cross-mode operation has been demonstrated, where the pump can be placed several GHz apart from the signal tone \cite{Zhao_2023}, and non-degenerate pump operation by pumping at two different frequencies, which is particularly stable. 
Furthermore, phase-sensitive amplification has been demonstrated, which is strongly linked to squeezing capabilities. 
Crucially, noise temperatures of better than \qty{1.5}{\K}, which is already more than twice as sensitive as the best HEMT amplifiers have been observed. 
An important observation is that we have been able to operate resonator amplifiers having high gain and low noise at physical temperatures of $\sim4\,\mathrm{K}$, which opens the door to multistage designs where the bulky and expensive HEMT amplifiers can be eliminated completely \cite{Zhao_2023,zhao2024_intrinsic_separation}.

Overall, superconducting kinetic inductance resonator amplifiers have been found to be easy to manufacture, and simple and robust to operate. They display flat gain profiles and appreciable saturation powers. Manufacturing approximately 50 amplifiers on a single wafer is a straightforward process, enabling the efficient production of large amplifier arrays required for a multi-channel CRES experiment \cite{withington2024}.

\subsection{SLUGs}

An alternative approach to low-noise microwave amplification is to use a superconducting quantum interference device (SQUID). Instead of employing the nonlinearity of a material or a Josephson junction to carry out wave mixing, a SQUID is used as a highly sensitive flux-to-voltage converter. The magnetic flux-periodic response of the SQUID, arising from its macroscopic quantum behaviour, implies that very high nonlinearity in the flux to voltage transfer function can be achieved.

A SQUID amplifier offers several attractive features, including robust fabrication, a good dynamic range, and compactness, which helps minimise the space it occupies within a cryostat.The superconducting properties also imply very low intrinsic noise \cite{Hao_2008}, allowing the standard quantum limit to be achieved.

SQUID amplifiers have been in development since the mid-80s \cite{Hilbert_1983, Hilbert_1985}. Early iterations routinely achieved gain in excess of 20 dB \cite{Muck_2010, Takami_1989, Welty_1993}. 
However the arrangement of the input circuit meant that operating frequencies were limited to hundreds of MHz. A subsequent development was to use a microstrip resonator to input the signal \cite{Muck_1998}, which had the effect of both enhancing the current at the amplifier input by the quality factor $Q$, and increasing the operating frequency. However, frequencies above a few GHz were still not practical, as the necessary shortening of the microstrip had the effect of decreasing the mutual coupling between the input and the SQUID.

To circumvent the frequency limits encountered with typical dc SQUID amplifiers the collaboration is developing a different variant of the SQUID $-$ the SLUG~\cite{Clarke_1966, Gallop_1974}. In a SLUG the signal to be amplified is coupled directly into the loop as a current, thereby enabling efficient coupling of frequencies of several GHz \cite{Ribeill_2011}.

\begin{figure}[t]
    \begin{adjustbox}{width=1.0\linewidth, center}
    \includegraphics{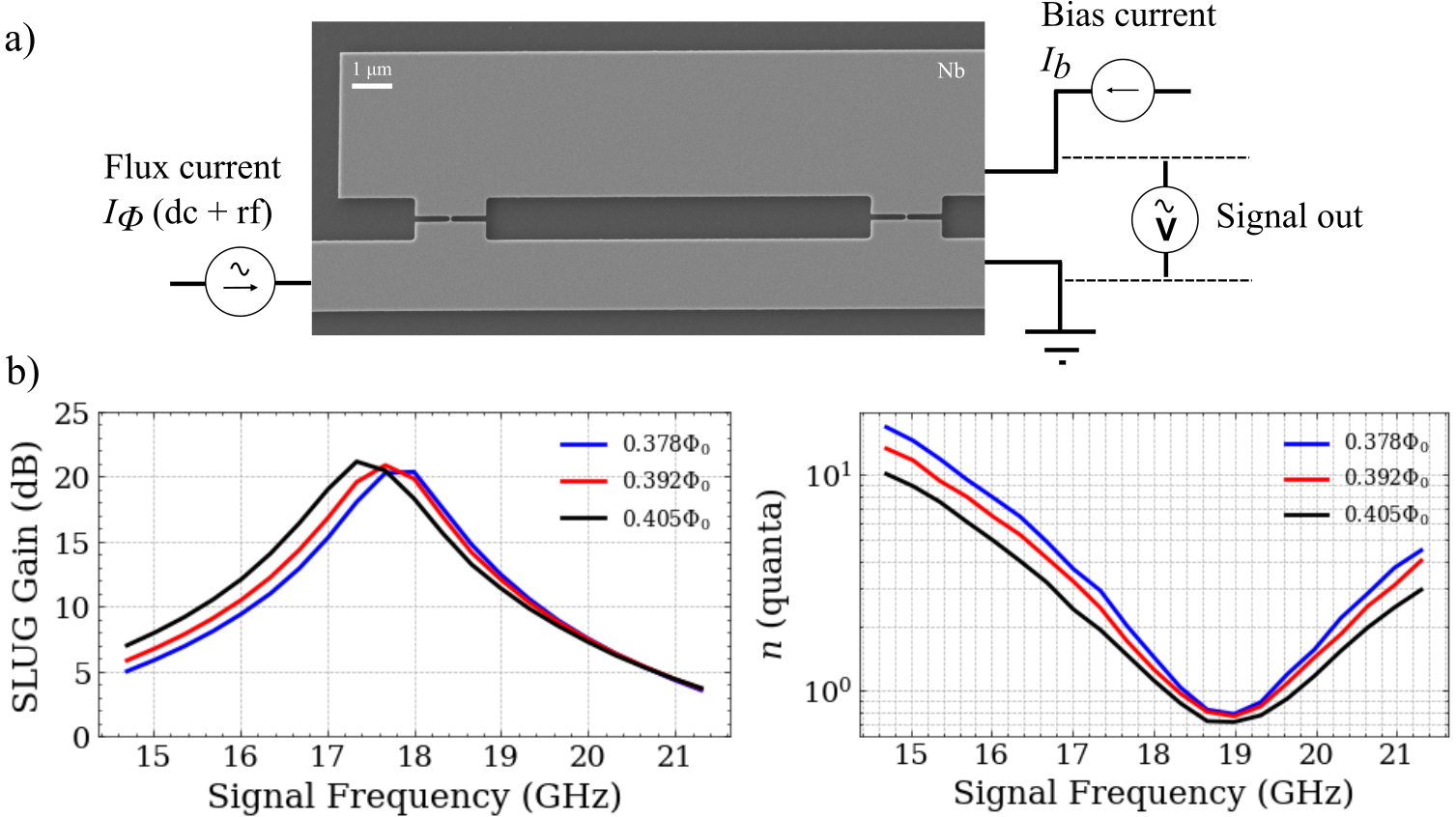}
    \end{adjustbox}
    \caption{(a) Typical planar nanobridge SLUG loop, and (b) simulated power gain and added noise for SLUG containing typical Nb nanobridges.}
    \label{fig:SLUG}
\end{figure}
Our goal is to develop a SLUG amplifier utilising niobium nanobridges as the Josephson elements. Nanobridges have been shown \cite{Polychroniou_2020} to have extremely high plasma frequency, due to their high critical current density and negligible capacitance, which makes them a natural choice for high frequency operation. The Nb nanobridge SLUG amplifier will also be entirely planar, fabricated from a single film of Nb, which is a significant structural simplification compared with existing technology \cite{Hover_2012}. The collaboration has developed a set of modelling tools for designing and optimising the SLUG. The SLUG loop can be designed to have a geometry which provides the appropriate inductance using 3D-MLSI and the gain and noise properties of the full amplifier can be modelled using a numerical model \cite{Ribeill_2011} which solves the time-dependence of the phase-difference across the nanobridges.

\Fref{fig:SLUG} shows a typical planar SLUG, and the results of simulation of a SLUG amplifier with input circuit designed to operate at 18 GHz. The circuit parameters used for this simulation were informed by dc measurements of protype Nb nanobridge SLUG circuits \cite{Chapman_2024}. We note that due to a difference in the optimal input impedance for gain and noise, the frequency of maximum gain and the frequency of minimum noise are not the same. It is, however, possible to achieve a gain of 20 dB and noise within a factor of 4 of the standard quantum limit at 18 GHz. We also note the flux-tuneability of the gain curve, which may prove to be a favourable property of the SLUG amplifier.

\section{Summary} 
\label{sec:summary}

The discovery of neutrino oscillations at the end of the last century remains the strongest evidence in support of physics beyond the Standard Model. 
Although the phenomenon of oscillations unequivocally proves that neutrinos have a non zero-mass, these experiments cannot measure the absolute value of the mass scale. The question of the absolute neutrino mass is crucial to understanding the origin of matter and the evolution of the early Universe. 
One promising, and the most model independent, approach to measuring this last unknown mass parameter of the Standard Model is through the precise measurement of the $\beta$-decay spectrum of atomic tritium. 
It is highly complementary to other neutrino mass probes, namely observations of the structure of the Universe and neutrinoless double $\beta$-decay.
It is a controlled, terrestrial experiment and, as a kinematic measurement, is largely model agnostic, independent of cosmological model assumptions and whether neutrinos are Dirac or Majorana fermions.

This White Paper outlines the QTNM project's innovative approach to tackle the neutrino mass puzzle by leveraging cutting-edge experimental methods in AMO physics and quantum technologies. 
These include using atoms in superpositions of Rydberg states as advanced sensors for magnetometry and electrometry and quantum-noise-limited amplifiers, originally developed for applications in astronomy, metrology, and communications. 
The proposed approach surpasses the capabilities of the state-of-the-art KATRIN experiment, and requires the development of a high phase-space density atomic tritium source, and an electron spectrometer capable of measuring electron energies near the 18.6 keV endpoint of the tritium $\beta$-decay spectrum with precision below 0.1 eV.

A large-scale experimental apparatus capable of performing the ultimate absolute neutrino mass measurement will need to handle significant quantities of tritium, creating the potential to host such an experiment at a national facility. 
QTNM’s strategy is to collaborate with leading facilities, including CCFE and the TLK, and international projects such as Project 8, KATRIN, and PTOLEMY, to build and operate a scalable apparatus with progressively increasing sensitivity to the absolute neutrino mass. 
The scientific milestones include sensitivities of around 0.1~eV/$c^2$ (degenerate neutrino masses), 0.05~eV/$c^2$ (inverted ordering of neutrino masses), and ultimately 0.01~eV/$c^2$, approaching the lower bound from neutrino oscillation experiments in case of the normal ordering of neutrino masses.

Beyond neutrino mass measurement, this approach will have the potential to explore beyond-the-Standard-Model physics, such as non-standard neutrino interactions, Lorentz-invariance violation, sterile neutrinos and exotic light bosons, and serve as a unique testbed for advanced quantum technologies, with applications extending beyond fundamental physics.

  %\begin{ack}
  \ack
  This work was supported by the UK Science and Technology Facilities Research Council (STFC) through the Quantum Technologies for Neutrino Mass (QTNM) project (grant no. ST/T006439/1). 
  NM is grateful to the STFC for their support through an Ernest Rutherford Fellowship (ST/W003880/1).
  ES is grateful for the support of the Erasmus+ Programme of the European Union.
  %\end{ack}

%\appendix
%\section{Sensitivity to Parameter Changes}
%
%\begin{figure}[t!]
%    \centering	
%    \includegraphics[width=0.48\textwidth]{Figures/sensitivity-var-%exposure}
%    \includegraphics[width=0.48\textwidth]{Figures/sensitivity-var-%baseBackground}\\
%    \includegraphics[width=0.48\textwidth]{Figures/sensitivity-var-molecularAdmixture}
%    \includegraphics[width=0.48\textwidth]{Figures/sensitivity-var-sigmaBoverB}\\
%    \includegraphics[width=0.48\textwidth]{Figures/sensitivity-var-usigmaB}
%   \includegraphics[width=0.48\textwidth]{Figures/sensitivity-var-usigmaf}
%    \caption{As Fig.~4, but showing the change in sensitivity when varying the indicated experimental parameter within a factor of $(0.1, 0.32,1,3.2,10)$.}
%\end{figure} 
%

% \section*{References} 
% \bibliography{bibliography.bib}

\printbibliography
% \bibliography{bibliography}

\end{document}